\documentclass[preprint,superscriptaddress,nofootinbib]{revtex4}

\usepackage{graphicx, hyperref, amsmath, amssymb, slashed, color, bbm}

\newcommand{\nc}{\newcommand}

\nc{\beq}{\begin{equation}}
\nc{\eeq}{\end{equation}}
\nc{\barray}{\begin{eqnarray}}
\nc{\earray}{\end{eqnarray}}
\nc{\barrayn}{\begin{eqnarray*}}
\nc{\earrayn}{\end{eqnarray*}}
\nc{\bcenter}{\begin{center}}
\nc{\ecenter}{\end{center}}
\nc{\ket}[1]{| #1 \rangle}
\nc{\bra}[1]{\langle #1 |}
\nc{\0}{\ket{0}}
\nc{\mc}{\mathcal}
\nc{\er}[1]{(\ref{eq:#1})}
\nc{\onehalf}{\frac{1}{2}}
\nc{\partialbar}{\bar{\partial}}
\nc{\psit}{\widetilde{\psi}}
\nc{\Tr}{\mbox{Tr}}
\nc{\hc}{\mbox{H.c.}}
\nc{\ev}{\;\mathrm{eV}}
\nc{\mev}{\;\mathrm{MeV}}
\nc{\gev}{\;\mathrm{GeV}}
\nc{\tev}{\;\mathrm{TeV}}

\def\chii0{\chi_i^0}
\def\chij0{\chi_j^0}

\newcommand{\gsim}{\lower.7ex\hbox{$\;\stackrel{\textstyle>}{\sim}\;$}}
\newcommand{\lsim}{\lower.7ex\hbox{$\;\stackrel{\textstyle<}{\sim}\;$}}
\nc{\ttbar}{t\bar t}
\nc{\TAFB}{A_{FB}^t}
\nc{\lepAFB}{A_{FB}^\ell}
\nc{\TAC}{A_{C}^t}
\nc{\Lag}{\mathcal{L}}
\nc{\Proj}{\mathcal{P}}

\def\Re{\text{Re}}
\def\Im{\text{Im}}

\def\sign{\text{sign}}

\begin{document}

\title{Open windows for a light axigluon explanation of $\TAFB$}

\author{Moira Gresham}
\affiliation{Whitman College, Walla Walla, WA 99362}
\author {Jessie Shelton}
\affiliation{Center for the Fundamental Laws of Nature, Harvard University, Cambridge, MA 02138}
\author{Kathryn M. Zurek}
\affiliation{Michigan Center for Theoretical Physics, University of Michigan, Ann Arbor, MI 48109}
\affiliation{ School of Natural Sciences, Institute for Advanced Study, Princeton, NJ 08540}

\begin{abstract}
  The top forward-backward asymmetry ($\TAFB$) measured at the
  Tevatron remains one of the most puzzling outstanding collider
  anomalies.  After two years of LHC running, however, few models for
  $\TAFB$ remain consistent with LHC data.  In this paper we take a
  detailed look at the most promising surviving class of models,
  namely light ($m_{G'} \lesssim 450$ GeV), broad axigluons.  We show
  which models simultaneously satisfy constraints from Tevatron and
  LHC top measurements, hadronic resonance searches, and LEP precision
  electroweak (PEW) observables. We consider three flavor structures:
  flavor-universal; down-type nonuniversal, designed to ease
  constraints from LHC charge asymmetry measurements; and top-type
  nonuniversal, designed to ameliorate constraints from PEW.  We
  compute contributions to the PEW observables from states in the
  minimal UV completion of the axigluon model and demonstrate that new
  heavy fermions make the constraints universally more stringent,
  while related contributions from new scalars are much smaller, but
  act to relax the constraints.  Paired dijet searches from ATLAS and
  CMS rule out all narrow axiglue models, while the LHC charge
  asymmetry measurement is less constraining than expected due to the
  high central value measured by ATLAS.  Excepting the tension with
  the CMS charge asymmetry measurement, a broad axigluon is consistent
  with all data over the entire mass range we consider ($50 \mbox{
    GeV} \lesssim m_{G'} \lesssim 450 \mbox{ GeV}$) in the
  flavor-universal and top-type nonuniversal models, while it is
  consistent for $m_{G'} \gtrsim 200 \mbox{ GeV}$ in the down-type
  non-universal model.  The LHC charge asymmetry remains the best
  avenue for excluding, or observing, these models.
\end{abstract}

\maketitle

\section {Introduction}

The anomalously large measurement of the top forward-backward
asymmetry $\TAFB$ at the Tevatron is one of the most significant and
puzzling outstanding collider anomalies. The CDF and D0 collaborations
have independently measured inclusive asymmetries approximately $2
\sigma$ above the Standard Model (SM) expectation; the most recent
measurements are detailed in Table~\ref{table: afb measurements}
\cite{Aaltonen:2012it, Abazov:2011rq}.  In addition, both experiments
have observed more significant discrepancies between measurement and
SM predictions in subsystems of the $t\bar t$ events.  Interest in the
$\TAFB$ exploded after CDF's 5.3 fb$^{-1}$ measurement
\cite{Aaltonen:2011kc} of a $\TAFB = 0.475 \pm 0.114$ asymmetry in
events with $M_{\ttbar} > 450 \text{GeV}$, which was 3.4$\sigma$ above
the SM prediction at the time.  In the updated measurement using the
full CDF data set, the high-mass excess has been mitigated, but still
grows very steeply with center of mass energy, and is 2.3 $\sigma$
above the SM expectation \cite{Aaltonen:2012it}.  Unfortunately D0
does not unfold their differential $\TAFB$ measurement, so it is not
possible to directly compare their results in the high invariant mass
range to those of CDF. D0 does, on the other hand, measure the lepton
asymmetry in $t\bar t$ events, which provides a clean and
theoretically sensitive cross check of the parent top asymmetry
\cite{Bernreuther:2010ny, Krohn:2011tw}.  D0 finds, in 5.4 fb$^{-1}$,
at production level $A^{\ell}_{FB} = 15.2\pm 4.0 \%$, which is more
than 3$\sigma$ above the {\tt MC@NLO} prediction of $A^{\ell}_{FB,SM}
= 2.1\pm 0.1 \%$ \cite{Abazov:2011rq}.  However, the significance of
this result has also been reduced with the addition of more data.
Combining with measurement of the (single) lepton asymmetry in
dileptonic top events, and including EW contributions in the SM
prediction, the updated result for the D0 single lepton asymmetry is
reduced to $A^{\ell}_{FB} =11.8 \pm 3.2\%$, a 2.2$\sigma$ discrepancy
with the SM \cite{:2012bfa}.  Meanwhile, CDF finds a $2\sigma$ excess
from the SM in the background-subtracted $\lepAFB = 6.6\pm 2.5\%$ with
a SM prediction of $1.6\%$ \cite{cdf-afb}.

While the deviation from SM predictions for the inclusive top
forward-backward asymmetry does not have high significance, the
consistency of the excess both across time and across experiments is a
possible indication of a non-statistical origin for the asymmetry.
The mystery is deepened by the excellent agreement of other top
properties with the predictions of the SM, and in particular by the
consistency of the $\ttbar$ production cross-section (both inclusive
and differential) between theory and experiment.

\begin{table}
\begin{tabular}{c c l}
\hline
$M_{\ttbar} $ & $\TAFB$  & Measurement/Prediction at Parton Level\\
\hline
inclusive & ~~$0.164 \pm 0.045$~~ & CDF \cite{Aaltonen:2012it}\\ 
& $ 0.196 \pm 0.065$  & D0  \cite{Abazov:2011rq} \\
 & $0.066 \pm 0.020$ & POWHEG SM prediction after applying EW corrections  \cite{Aaltonen:2012it} \\
\hline
$> 450$ GeV & $0.295 \pm 0.058 \pm 0.031$ & CDF  \cite{Aaltonen:2012it}\\
&  $0.100 \pm 0.030$ & POWHEG SM prediction after applying EW corrections \cite{Aaltonen:2012it}\\
\hline
 & $\lepAFB$  & \\
 \hline
 & $0.152 \pm 0.04$ & D0 \cite{Abazov:2011rq} \\
 & $0.118 \pm 0.032$ & D0, dileptonic \& semileptonic, combined \cite{:2012bfa} \\
 & $0.047 \pm 0.001$ & (D0) MC{@}NLO plus EW  \cite{:2012bfa} \\
 & $ 0.066 \pm 0.025 $ & CDF, background subtracted  \cite{cdf-afb}\\
 & $ 0.016 $ & (CDF) NLO (QCD + EW)  \cite{cdf-afb}\\
 \hline
\end{tabular}
\caption{Recent measurements of $\TAFB$ and $\lepAFB$ along with SM predictions.}\label{table: afb measurements}
\end{table}

Very many new physics models have been proposed to explain the
anomalously large top asymmetry.  Most have addressed the tension
between the discrepant $A_{FB} $ and the well-behaved cross-section by
deferring predicted deviations in the spectrum to partonic center of
mass energies beyond the Tevatron's reach.  For heavy $s$-channel
particles such as axigluons
\cite{Frampton:2009rk,Chivukula:2010fk,Bai:2011ed,Haisch:2011up,Gresham:2011pa,Barcelo:2011vk,
  Alvarez:2011hi}, which have large masses $m\gsim$TeV as well as
broad natural widths $\Gamma\gsim 0.3\, m$, significant deviations
from SM predictions for the dijet and top pair spectra are inevitable
at and above a TeV, as center of mass energies begin to approach the
axigluon pole.  Meanwhile models that generate the asymmetry via the
$t (u)$-channel exchanges of flavor-violating/carrying vectors
(scalars)
\cite{Jung:2009jz,Shu:2009xf,Gresham:2011dg,Shelton:2011hq,Jung:2011zv,Jung:2011id}
typically involve mediators significantly lighter than a TeV with
large, flavor off-diagonal couplings.  These models attain reasonable
agreement with Tevatron top cross-sections by arranging a cancellation
between interference and new-physics terms at Tevatron energies.  This
cancellation no longer holds at LHC energies, so while these models do
avoid producing a dijet or $t\bar t$ resonance, the high-$m_{t
  \bar{t}}$ tail in top pair production is strongly enhanced
(suppressed) \cite{Gresham:2011fx,Delaunay:2011gv,AguilarSaavedra:2011vw}.  Models with sufficiently light and weakly coupled
mediators $M$ can avoid over (under)-producing $t\bar t + X$; however,
the large single top production in these models, $t + M\to t + j j$,
contributes at unacceptable levels to top pair cross-section analyses.
Moreover, processes in which the mediators are directly produced
on-shell in association with the top quark lead to many distinctive
and charge-asymmetric processes that contribute to single top and top
pair final states \cite{Craig:2011an,Knapen:2011hu}.  Top-jet
resonances arise in these models \cite{Gresham:2011dg}, which have
been searched for and excluded over much of the parameter space
\cite{Chatrchyan:2012su,ATLAStj}. Top cross-section measurements at
the LHC thus exclude these classes of models when all contributions to
top-pair-like final states are taken into account, unless additional
BSM decay modes for the mediator are introduced to hide it
\cite{Alvarez:2012ca,Drobnak:2012rb}.  $t
(u)$-channel models are also strongly constrained by low-energy Atomic
Parity Violation (APV) measurements \cite{Gresham:2012wc} and the
failure of the LHC experiments to observe large charge asymmetries
\cite{AguilarSaavedra:2011vw,ATLAS:2012an,:2012xv} or deviations from SM predictions in top
polarization and spin correlations \cite{Fajfer:2012si}.

As the LHC has thus far failed to find significant deviations from
standard model predictions for single top or $\ttbar$ processes, using
heavy new states to explain the top forward-backward asymmetry is now
increasingly disfavored \cite{Delaunay:2012kf}.  Only small regions of
parameter space remain for heavy axigluons with the top quark coupling much
larger than light quark coupling to evade dijet constraints.

As an alternative approach, new physics explanations for the top
forward-backward asymmetry can instead invoke {\it light} axigluons
\cite{Tavares:2011zg,AguilarSaavedra:2011ci,Krnjaic:2011ub,Drobnak:2012cz},
which can be more weakly coupled and therefore lead to much smaller
deviations from SM predictions for top properties.  Here by ``light''
axigluons, we mean models where the light quark and top quark axial
couplings have the same sign, $\sign (g_A^q) =\sign (g_A^t)$.  In
order to generate the observed sign for the inclusive forward-backward
asymmetry, these axigluons must therefore be not much heavier than
$\sim 2 m_t$. These light axigluons would be copiously produced at
current and past colliders, and require model building to be
``hidden'' from discovery under large QCD backgrounds.

We will examine the existing constraints on light, hidden axigluons
and related particles.  Direct collider searches for narrow resonances
decaying to dijets entirely eliminate narrow axigluons above 100 GeV.
Below the $Z$ pole axigluons run into constraints from the running of
$\alpha_s$, and are excluded for masses below approximately 50 GeV
\cite{Kaplan:2008pt}.  For sufficiently broad and weakly coupled
axigluons, it is possible to avoid discovery in direct collider
searches.  In these cases the most important constraints come from two
indirect measurements.  First, the one-loop axigluon corrections to
the $Z \rightarrow q \bar q$ coupling constrains light axigluon models
through the LEP precision electroweak (PEW) measurements of the
hadronic $Z$ width and the total hadronic cross-section at the $Z$
pole \cite{Haisch:2011up}.  Second, the non-observation of a large
charge asymmetry at the LHC is also becoming constraining for light
axigluons \cite{Drobnak:2012cz}.  These indirect constraints are
highly sensitive to the flavor structure of the axigluon-quark
couplings. As we will see, the constraints from PEW observables and
from the LHC charge asymmetry measurements make competing demands on
the flavor structure, which significantly limit the allowed parameter
space.

We will discuss three flavor structures. First, we consider
flavor-universal axigluons. Second, we consider axigluons where the
coupling to right-handed down-type quarks is enhanced, a choice which
helps to reconcile LHC and Tevatron charge asymmetry measurements
\cite{Drobnak:2012cz}, but exacerbates the tensions with PEW
observables.  Third, we consider axigluons with an enhanced coupling
to top quarks, a choice motivated by minimal flavor violation-type
models and models with a special role for the third generation, which
alleviates the tension with the PEW observables but does not help with
the LHC charge asymmetry measurement.  These models also can run into
difficulty with the lepton asymmetry measured at the Tevatron.

The outline of this paper is as follows.  In the next section we
present an overview of the Tevatron forward-backward and LHC
forward-central charge asymmetry measurements and identify two
interesting non-minimal flavor structures that are safe from
low-energy precision measurements.  We then summarize existing
constraints on light axigluons from top pair production that are
largely independent of the axigluon width: the forward-backward and
forward-central charge asymmetries in section \ref{sec:tafb}, the
lepton asymmetry in section \ref{sec:lafb}, and total cross-section in
\ref{sec:xsec}.  We discuss constraints from direct collider searches,
in particular from paired dijets, in section
\ref{sec:colliders}. Precision EW constraints for both the axigluon
alone and for various extensions of the broad axigluon model are
considered in section \ref{sec: pew}.  Finally we assemble the
constraints and perform a global fit, identifying surviving regions in
parameter space.  We refer the reader to Figs.~\ref{fig: summary
  curves: flavor universal},~\ref{fig: summary curves: flavor
  nonuniversal} for a summary of the open windows for a light axigluon
explanation of $\TAFB$.  While this work was in preparation,
\cite{Gross:2012bz} appeared, which has overlap with this work.

\section{Models and Conventions}
\label{sec:model}

In this section we define a minimal reference Lagrangian for a light
axigluon and discuss the three flavor structures we will focus on.
We describe the axigluon as arising from a spontaneous breaking of
$SU(3)_1\times SU(3)_2\to SU(3)_c$.  This is the minimal
renormalizable realization of a massive vector octet, and gives the
Lagrangian for the axigluon $G'$ and SM gluon $G$:
\beq
\label{eq:UVlag}
\Lag = - \frac{1}{4} (D_\mu {G'}_{\nu} )^ a(D_\mu {G'}_{\nu} )^ a
-{g_s \over 2} \chi \, f^{abc}\, G^{\mu\nu a} {G'}_{\mu}^b {G'}_\nu^c 
\eeq
where 
\beq
(D_\mu {G'}_{\nu} )^ a=\partial_\mu  {G'}_\nu + g_s f ^ {a bc} G^b_\mu  {G'}_\nu ^ c - 
        (\mu\leftrightarrow \nu) .
\eeq
and the coefficient of the second term in Eq.~\ref{eq:UVlag}, which in the low-energy theory
is undetermined, is fixed in the UV completion to be $\chi = 1$.

Axigluon couplings to quarks,
\beq
\Lag = - \sum_{i=1}^3\left( g_{L, i} \bar Q_L^i  \slashed{G'} Q_L^i + g_{R,i}^D \bar d_R^i\slashed{G'} d_R^i     + g^U_{R,i}  \bar{u}_R^i \slashed{G'} u_R^i \right),
\eeq
on the other hand, are model-dependent.  In general, axigluon-quark
couplings $g_i$ smaller than $g_s$ are necessary for light axigluons
to give a good fit to the Tevatron data.  Since simple embeddings of
the quark generations into the minimal UV group $SU(3)_1\times
SU(3)_2$ give axial couplings bounded from below by $g_s$, the small
couplings needed to explain the Tevatron data are challenging to
obtain without invoking new degrees of freedom \cite{Tavares:2011zg,
  Cvetic:2012kv}, as summarized in appendix~\ref{sec:UV}.  Our
standpoint here will be purely phenomenological, using a simple
low-energy Lagrangian with freely-adjustable couplings between
axigluons and quarks. However, as the structure of the minimal UV
completion is sharply defined, and as some of the additional degrees
of freedom could provide natural additional decay channels for the
axigluon, we will also consider contributions to PEW observables from
additional heavy degrees of freedom in Section~\ref{sec: pew}.

We concentrate on three patterns for the quark-quark-light axigluon
couplings that are compatible with flavor constraints without
fine-tuned alignment of mass and flavor bases\footnote{For a
    systematic account of flavor-symmetric models in the context of
    $\TAFB$, see \cite{Grinstein:2011yv,Grinstein:2011dz}.}. We consider the
Lagrangian,
\beq
\Lag = - \left( g_L \bar{Q}_L^i \slashed{G'} Q_L^i +g^D_{R} \bar{d}_R^i \slashed{G'} d_R^i  + g^U_{R} \bar{u}_R^i \slashed{G'} u_R^i + \delta^t_{R}  \bar{t}_R \slashed{G'} t_R  \right).
\eeq
with the four parameters $g_L, g^D_R, g^U_R, \delta^t_R$ occurring in
the combinations given in Table~\ref{axigluemodels}.  This defines
three flavor scenarios: (i) {\it flavor universal}
\cite{Tavares:2011zg, Krnjaic:2011ub} (ii) {\it down-type
  non-universal} \cite{Drobnak:2012cz, AguilarSaavedra:2012va} and
(iii) {\it top non-universal}.  The down-type non-universal scenario
is preferred by CMS LHC charge asymmetry measurements, and the top
non-universal scenario is preferred by PEW measurements.  In all three
scenarios, couplings to up quarks are chosen to be purely axial ($g_A
= {1\over 2} (g_R - g_L) \neq 0, \; g_V = {1\over 2} (g_R + g_L) =
0$), since this is the choice that maximally affects top charge
asymmetries while minimizing the effect on charge-symmetric top
observables.  We consider enhancement of only the RH top coupling (not
RH bottom or LH top-bottom doublet) because it is motivated by minimal
flavor violation and---more importantly---because constraints are
weakest: constraints on models with $b$ coupling enhancement as well
would only increase.  Axigluons with mass below the top quark require
very moderate couplings in order to generate a charge asymmetry
commensurate with the measured Tevatron values and are therefore
typically narrow if their only allowed decays are to quarks. Since
dijet resonance constraints rule out most such models, we consider
both narrow and broad axigluons. For concreteness we will take 20\% as
a benchmark ``broad'' width and the natural width to light quarks as a
``narrow'' width.

\begin{table}
\begin{tabular}{c c}
  Scenario Name & Couplings  \\
  \hline
  Flavor universal  &  ${g_R^U} = g_R^D =  - g_L \equiv g_A; \; \delta^t_R=0 $     \\
  Down-type non-universal  &  ${g_R^U}  =  - g_L \neq g_R^D; \; \delta^t_R=0 $     \\
  Top non-universal  &  ${g_R^U} = g_R^D =  - g_L \equiv g_A; \; \delta^t_R = g^t_R - g^U_R \neq 0 $ \\
  \hline
  \end{tabular}
  \caption{Definitions of the axiglue scenarios we consider.}
  \label{axigluemodels}
\end{table}

An axigluon with large enough couplings to light quarks and the top
quark to generate the asymmetry at the Tevatron must satisfy several
non-trivial constraints.  We outline the constraints we will detail
below for the three classes of axiglue models we consider.
\paragraph*{(1) Flavor Universal:}
\begin{itemize}
\item LHC charge asymmetry. For narrow and broad axigluons, the
  Tevatron and LHC charge asymmetry as measured by CMS are in mild
  tension for the entire mass range. However, the ATLAS charge
  asymmetry is perfectly commensurate with Tevatron $\TAFB$.
\item Precision electroweak (PEW) constraints, dominantly from
  one-loop corrections to the $Z$-$q$-$\bar q$ vertex. These
  constraints strongly disfavor a sub-100 GeV narrow or broad
  axigluon.
\item Single and paired dijet constraints. Narrow axigluons are
  strongly disfavored by single dijet resonance searches at hadron
  colliders in all but the sub-$m_Z$ mass range.  Paired dijet
  searches also rule out the entire narrow resonance window from
  constraints on production of two axigluons that decay to pairs of
  jets.
\end{itemize}

As we will show, these combined constraints leave a strip of parameter
space open for a light flavor universal axigluon heavier than $m_Z$.
A lower charge asymmetry measurement from ATLAS would strengthen the
constraints on these models considerably.  Individual constraints can
be partially or fully alleviated in flavor non-universal models.
Constraints from a low LHC charge asymmetry can be alleviated by
increasing couplings to down-type quarks \cite{Kamenik:2011wt}.
Precision electroweak constraints can be relaxed by allowing the light
quark couplings to be small by simultaneously increasing the top quark
couplings. Paired dijet constraints still eliminate all flavor
non-universal models with a narrow axigluon; broad axigluons survive.

\paragraph*{(2) RH Up-Down Flavor non-Universal Axigluons:}
\begin{itemize}
\item By taking the coupling to the down quark larger than to the
  up-type quarks, constraints from the CMS LHC charge asymmetry can be
  eliminated.
\item Precision electroweak constraints are particularly stringent in
  this case, requiring the axigluon to be heavier than 200 GeV.
\item Even though alleviating the tension with
  the CMS LHC charge asymmetry requires RH down-type quark couplings
  of order $g_s$, the consequent increase in the Tevatron top pair
  cross-section is still small.
\item As for all flavor choices, paired dijet constraints eliminate
  narrow axigluons over the entire mass range.  For this case, broad
  axiglue are also constrained by UA1 dijets.
\end{itemize}

\paragraph*{(3) RH Top non-Universal Axigluons:}
\begin{itemize}
\item These models do nothing to alleviate the CMS LHC charge asymmetry
  constraint.
\item By taking the coupling to RH top much larger than that to the
  light quarks, precision electroweak
  constraints are alleviated.
\item  These models can over-predict the Tevatron lepton asymmetry, particularly
 for axigluons below the $2m_t$ threshold.
\end{itemize}

In the following sections we discuss in depth the observables and
constraints for each of the above-mentioned scenarios.

\section{Top pair observables}\label{sec: ttbar observables}

We begin by identifying the parameter ranges that produce sufficiently
large asymmetries at the Tevatron and illuminate any tension with
other $t \bar{t}$ observables such as the LHC charge asymmetry and the
$t \bar t$ cross-section.  We also discuss the Tevatron lepton
asymmetry constraints on these scenarios, which can be important for
large non-universal axigluon couplings to $t_R$.

\subsection{Tevatron $\TAFB$ and LHC $\TAC$}
\label{sec:tafb}

The charge asymmetry at the LHC $\TAC$ is highly correlated with the
forward-backward asymmetry $\TAFB$ at the Tevatron, and provides one
of the most direct cross-checks of the Tevatron measurement. The
current situation for $\TAC$ at the LHC is rather unclear. $\TAC$ as
measured by ATLAS in dileptonic events, $\TAC =
0.057\pm0.024\pm0.015$, differs by more than a standard deviation from
the CMS measurement using semileptonic events, $\TAC = 0.004\pm0.01\pm
0.012$\footnote{The recent dileptonic measurement from CMS
  \cite{CMS-PAS-TOP-12-010} has very large statistical uncertainties,
  and is not included.}. The two collaborations are more consistent
if the semileptonic result from ATLAS with $1/4$ the data is
included. For this reason, the $A_C$ constraint from the LHC is not as
strong as expected by this point from the data.

\begin{figure}
\begin{centering}
\begin{tabular}{c c}
Flavor universal  & Flavor universal \\
\includegraphics[width=0.48\textwidth]{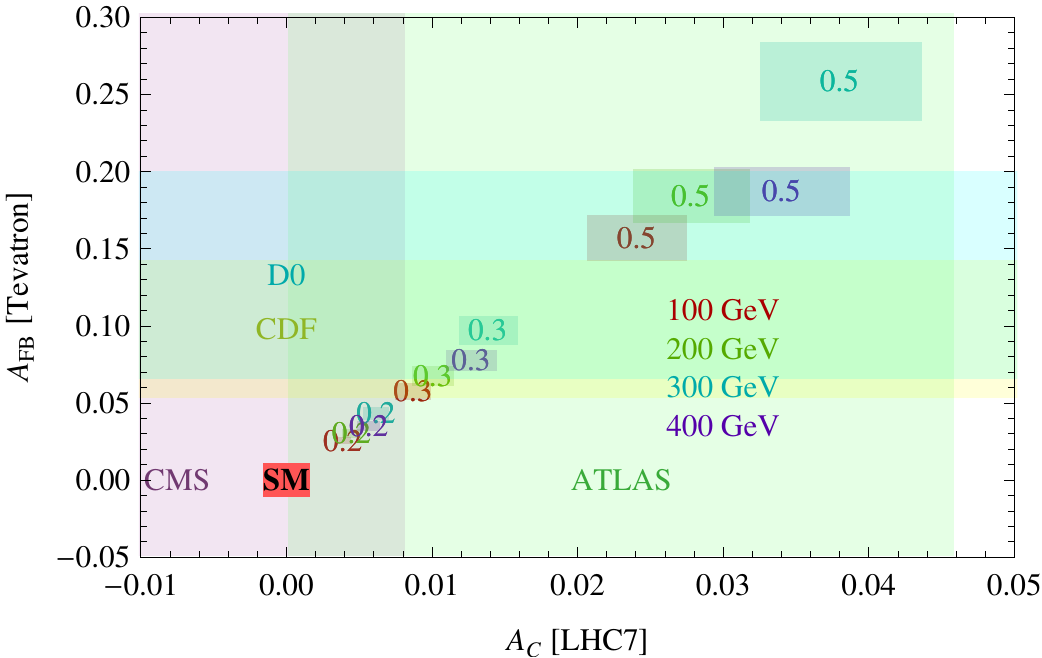} & \includegraphics[width=0.48\textwidth]{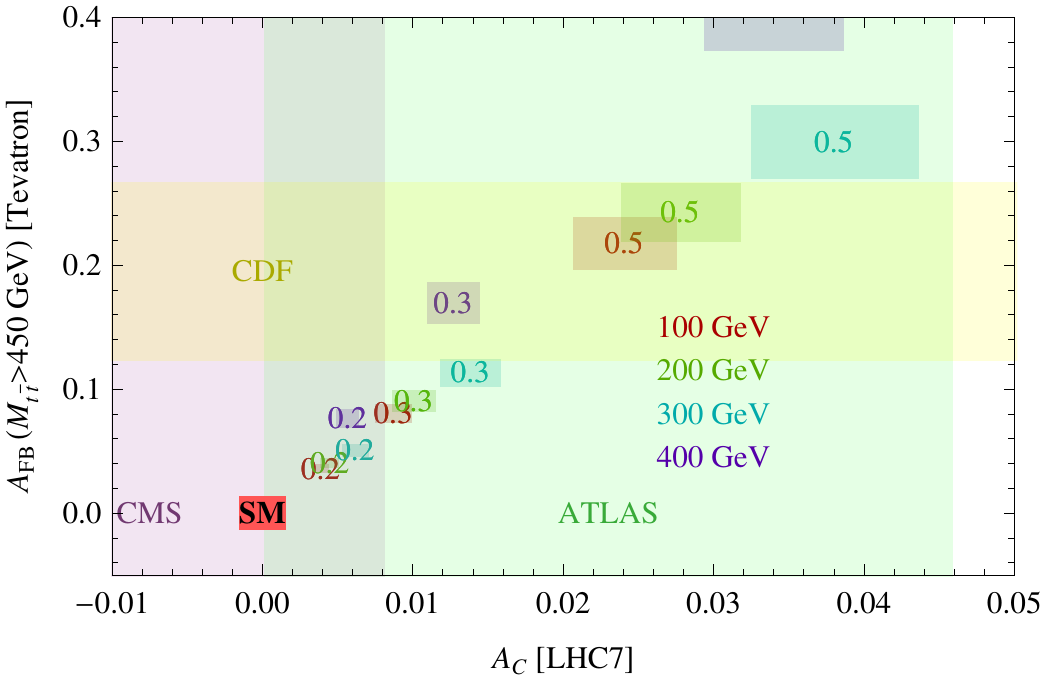}
\\
Down-type non-universal & Top non-universal \\
\includegraphics[width=0.48\textwidth]{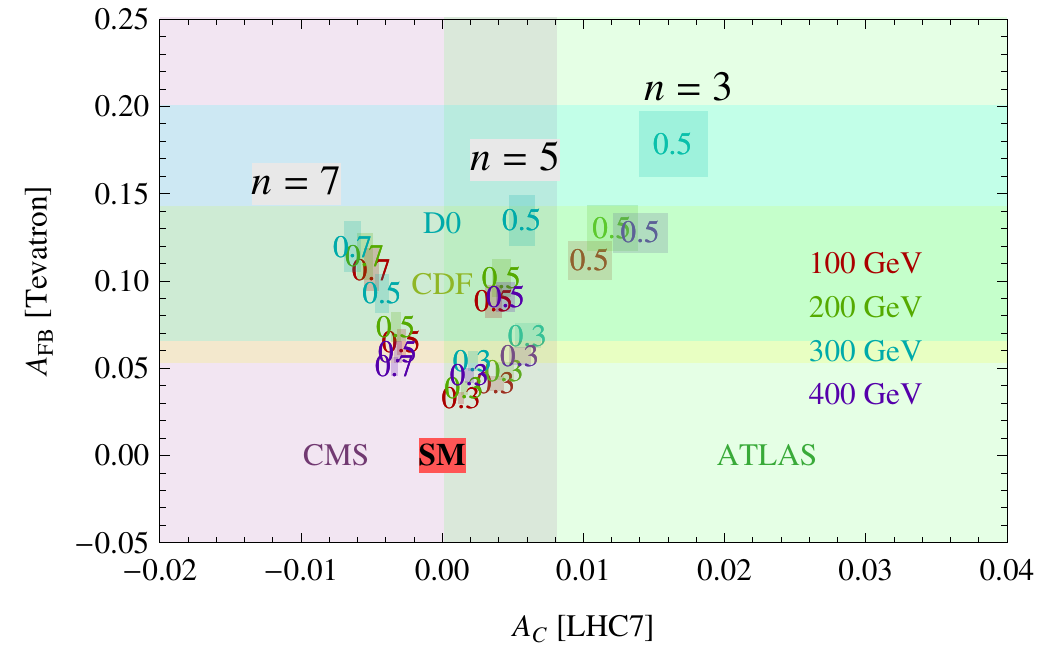}  & \includegraphics[width=0.48\textwidth]{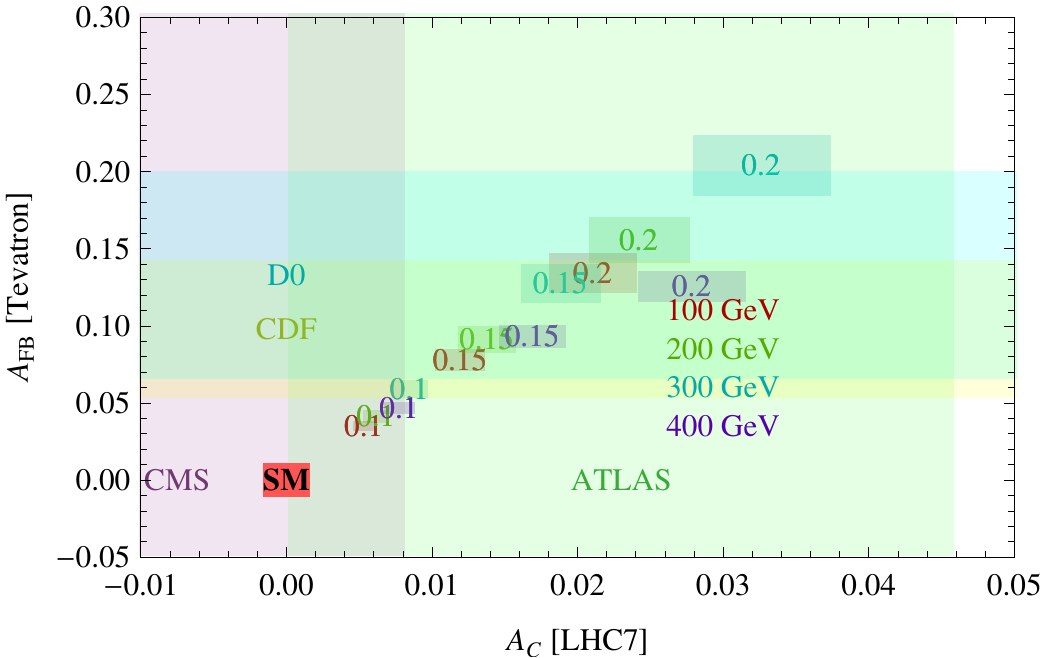} \\ 
\end{tabular}
\end{centering}
\caption{Asymmetries in $q \bar{q} \rightarrow t \bar{t}$ events ($t
  \bar{t} j$ events are a minor correction and not included here) for axigluon models with
  masses 100, 200, 300, and 400 GeV. Widths have been set to 20\%,
  though the asymmetries are insensitive to variations of width
  between 1\% and tens of percent for masses below $2 m_t$. Plot
  markers indicate $g_R^U = - g_L$ in units of $g_s$. Down-type universal models shown have $g_R^D=-n g_R^U$ and the top non-universal model shown has $g_R^t = 10 g_R^U$. Shaded
  rectangles behind plot markers show the variation in $\TAFB$ and
  $\TAC$ for factorization/renormalization scales varying from $m_t /
  2$ to $2 m_t$.  The most recent CDF (light yellow), D0 (cyan), CMS
  (light purple), and ATLAS (dileptonic + semileptonic combination,
  light green) 1$\sigma$ bands are shown. The bands are centered on
  the central value \emph{minus} the SM NLO expectation as reported by
  each collaboration.}\label{ud nonuniversal scan}
\end{figure}

Figs.~\ref{ud nonuniversal scan} and \ref{diff afb ac} show $\TAFB$
and $\TAC$ for light axigluon models in all three flavor
structures. The contributions to $\TAFB$ and $\TAC$ due to leading
order (LO) new physics are shown for various choices of parameters
alongside CDF and D0 bands corresponding to the measured asymmetry
minus the standard model expectation as reported by the collaboration,
with errors given by the experimental and SM prediction uncertainties
added in quadrature. We assume linear addition of SM and BSM
contributions to the asymmetry. Following CDF, we multiply D0's
reported QCD-only SM prediction by 1.26 to account for EW corrections
and include a $30\%$ error on the SM expectation.   We use QCD predictions as
 reported by the experiments, though there is some concern that these
 predictions are underestimates \cite{Manohar:2012rs}. Calculations are
semi-analytic. We used CTEQ5 parton distribution functions with $m_t =
173$ GeV and set the renormalization and factorization scales to
$m_t$; sensitivity to the renormalization/factorization scales was
checked by varying scales between $m_t/2$ and $2 m_t$.
Flavor-universal couplings are in tension with the CMS result, but not
the ATLAS result.  Down-type non-universal models can provide a better
fit to $\TAFB$ and a lower $\TAC$ \cite{Drobnak:2012cz}, while top
non-universal models do not alleviate the tension between $\TAFB$ and
$\TAC$.

\begin{figure}
\includegraphics[width=0.48\textwidth]{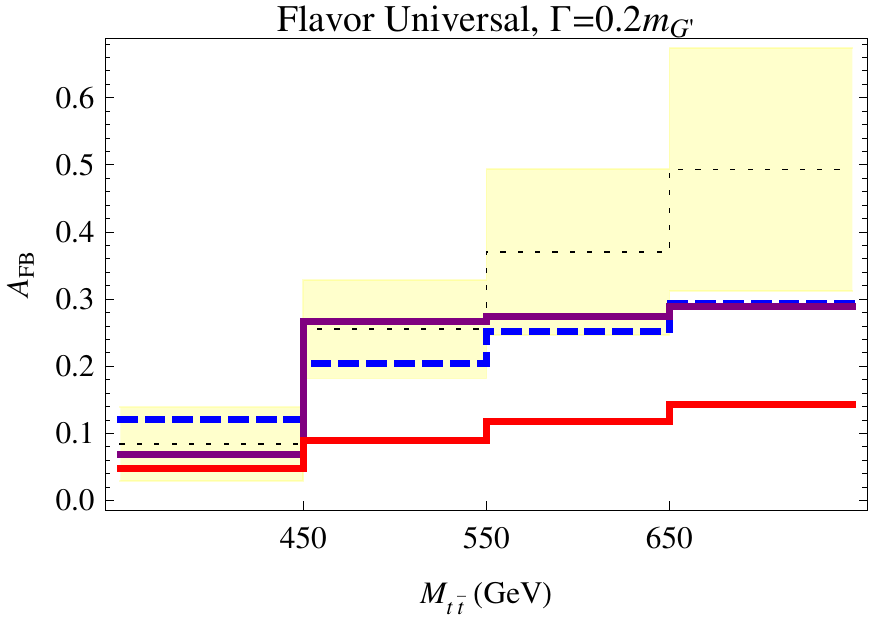} 
\includegraphics[width=0.48\textwidth]{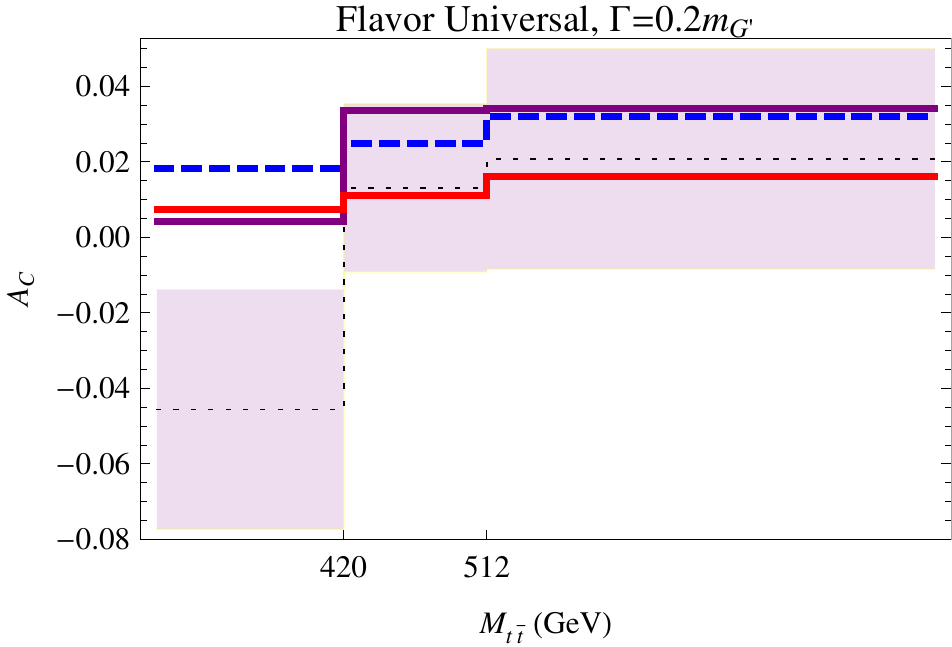} 
\caption{ Asymmetries binned according to $M_{t \bar{t}}$ for a
  200 (dashed blue) and 400 (solid purple) GeV axigluon with universal
  coupling of strength $0.35 g_s$ and $0.3 g_s$, respectively, to light and
  heavy quarks. Left: $A_\text{FB}$ shown alongside the most recent
  unfolded CDF measurement (light yellow indicates $1 \sigma$ band)
  and the SM NLO (QCD+EW) expectation (red). Right: $A_\text{C}$ shown
  alongside the most recent CMS measurement (light purple indicates $1
  \sigma$ band) and the SM NLO (no EW) expectation (red). }\label{diff
  afb ac}
\end{figure}

\subsection {Lepton Asymmetry}
\label{sec:lafb}

The forward-backward asymmetry of the charged lepton in semi-leptonic
top events, and the related asymmetry of the two oppositely-charged
leptons in dileptonic top events, is an interesting cross check of the
top forward-backward asymmetry. First, the lepton asymmetry $\lepAFB$, defined as
\begin{equation}
A_{FB}^\ell = \frac{N_\ell(Q \cdot \eta > 0) - N_\ell(Q \cdot \eta < 0)}{N_\ell(Q \cdot \eta > 0) + N_\ell(Q \cdot \eta < 0)},
\end{equation}
where $\eta$ is the rapidity of the lepton and $Q$ its charge, is
experimentally cleaner than the top asymmetry $\TAFB$, as it can be
measured without recourse to any top reconstruction procedure
\cite{Bernreuther:2010ny,Bernreuther:2012sx}.  Second, because the
lepton is highly sensitive to the potential existence of BSM angular
correlations in $\ttbar$ production, $\lepAFB$ provides {\it
  independent} information about the potential presence of BSM physics
in top pair production \cite{Krohn:2011tw}.

The size of the lepton asymmetry is determined by both (1) the
kinematics of the parent tops, and (2) the direction of the lepton in
the top rest frame.  Deviations from SM expectations for either the
kinematic distribution of top quarks or the angular distribution of
leptons in top decays will therefore alter the relationship between
$\TAFB$ and $\lepAFB$.  In particular, if the tops have some degree of
polarization, then nontrivial angular distributions of the top decay
products can substantially increase (for right-handed tops) or
decrease (for left-handed tops) the lepton asymmetry that arises from
kinematics alone.  Similarly, the presence of BSM spin correlations in
the top pair production amplitude induces non-SM-like dependence of
the lepton asymmetry on the center of mass energy
\cite{Falkowski:2011zr}.  Another possible mechanism to increase the
lepton asymmetry relative to the top asymmetry is to preferentially
produce top quarks that are hard and forward, such that the lepton and
top directions of flight as observed in the lab frame are more
correlated than in the SM.

Models with $t (u) $-channel mediators preferentially produce hard,
forward, right-polarized top quarks, and therefore predict a
significant enhancement of the lepton asymmetry, both relative to the
SM predictions for $\lepAFB$ and relative to $\TAFB$.  Axigluon models
produce more central top quarks, and (except in the non-universal top
scenarios) do not give rise to polarized tops, and consequently
predict smaller lepton asymmetries than do the $t(u)$-channel models.

\begin{table}
\begin{tabular}{l c c c c c}
Scenario   & $m_{G'}$(GeV)  & $g_{u_R}$($g_s$)  & $\lepAFB$ (incl.)  & $\lepAFB$ (D0 cuts)  & $\lepAFB$ (CDF cuts)   \\
  \hline

  Flavor universal  & 225 & 0.3 
                        & $2.9\pm 0.32$ & $2.2 \pm 0.51$ & $1.5 \pm 0.51$ \\
 
  Flavor universal  & 400 & 0.35 
                        &  $5.7\pm 0.41 $ &  $ 4.9 \pm 0.66 $ & $3.7\pm 0.74$  \\
  Down-type ($g_{d_R} = -5 g_{u_R}$)   &  350 & 0.4 
                        &$ 6.5\pm 0.32 $ &  $4.9 \pm 0.51 $& $3.9\pm 0.57$ \\
  \hline
  \end{tabular}
  \caption{Axigluon contribution to the parton-level lab frame lepton asymmetry for specific 
    benchmark points.  Asymmetry values quoted are percents. Errors on the predictions are from Monte Carlo statistics.  The 
    inclusive values are calculated using semi-leptonic $t \bar t$ events at parton level 
    with no cuts.  The D0 and CDF cuts applied to the inclusive asymmetry are taken 
    from \cite{Abazov:2011rq} and \cite{cdf-afb}, respectively.
    \label{table:leptonAFB}}
\end{table}

\begin{figure}
\begin{centering}
\includegraphics[width=0.7 \textwidth]{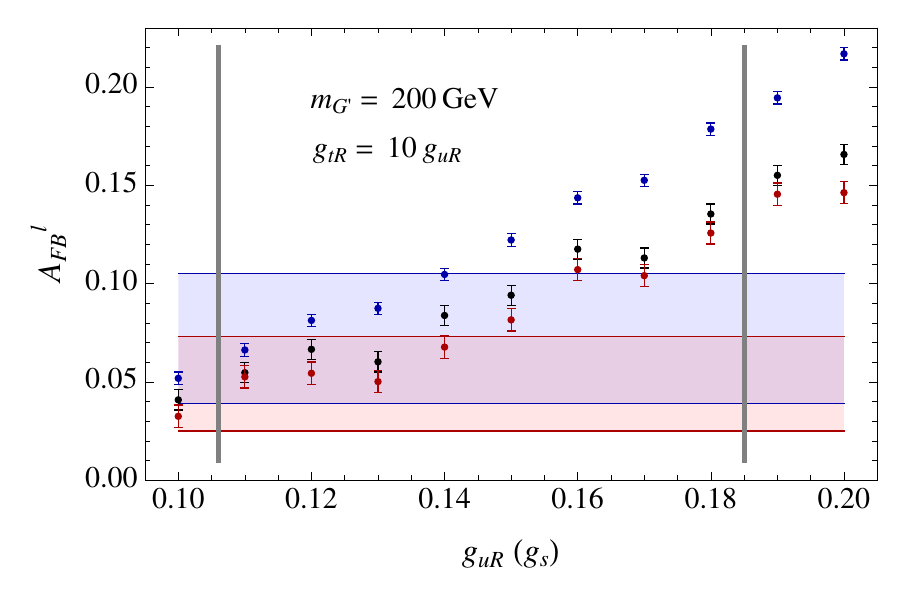} 
\end{centering}
\caption{ Contributions to the parton-level lab frame lepton asymmetry
  for a top non-universal axigluon with $m_{ G'} = 200 $ GeV and
  $g_{t_R}=10g_{u_R}$, for inclusive semi-leptonic $t\bar t$ events
  (blue), events passing semi-leptonic selection cuts after D0
  \cite{Abazov:2011rq} (black), and events passing selection cuts
  after CDF \cite{cdf-afb} (red).  Error bars show Monte Carlo
  statistical error. One-sigma preferred regions are shown in blue for
  D0's unfolded measurement (to be compared approximately to the
   inclusive events) and in red for CDF's reconstruction-level
  measurement (to be compared to the events with CDF's selection cuts)
  are shown in the shaded blue and red bands respectively.  Vertical
  lines indicate the upper and lower boundaries of the globally
  preferred region of Fig.~\ref{fig: summary curves: flavor
    nonuniversal}. \label{fig:lafb}}
\end{figure}

The lepton asymmetry as measured by D0 is 2.2$\sigma$ larger than SM
expectations.  This is large, but not sufficiently larger than the
corresponding excess in the inclusive top asymmetry as to decisively
point to BSM sources of top polarization.\footnote{Indeed, the lack of
  deviations in top polarization and related observables as observed
  at the LHC can constrain many new physics models for the Tevatron
  $\TAFB$ \cite{Fajfer:2012si}.}  In Table~\ref{table:leptonAFB} we
show for illustration the axigluon contributions to the lab-frame
$\lepAFB$ for some points that are characteristic of the parameter
spaces that will ultimately lie in the best-fit regions (see
Figs.~\ref{fig: summary curves: flavor universal}--\ref{fig: summary
  curves: flavor nonuniversal} below).  Results are shown at
parton-level, for the lepton asymmetry in semi-leptonic events, both
inclusive and those that pass selection cutsv as in
\cite{Abazov:2011rq,cdf-afb}.  The one-sigma allowed range for the BSM
contribution to the lepton asymmetry, as computed from D0's latest
measurement \cite{:2012bfa} and assuming linear addition of SM and BSM
contributions, is
\beq
\left. {\Delta\lepAFB}^{1\sigma}\right |_{\mathrm{parton}} = (3.9, 10.3) \% 
\eeq
while from CDF it is \cite{cdf-afb} 
\beq
\left. {\Delta\lepAFB}^{1\sigma}\right|_{\mathrm{sel.\,cuts}} = (2.5, 7.5) \% 
\eeq
Note the first number is at parton level after unfolding, and is
roughly what the inclusive asymmetries we show should be compared to.
We find that the lepton asymmetry generically favors slightly larger
couplings than does the top asymmetry in flavor-universal and
down-type nonuniversal axigluon models, as the $m=225$ GeV benchmark
in Table~\ref{table:leptonAFB} illustrates, but most of the global fit
preferred region is entirely consistent with the one-sigma range for
the lepton asymmetry.  By contrast, top non-universal models {\it
  overproduce} the lepton asymmetry over much of the global fit
preferred region, as can be seen in Fig.~\ref{fig:lafb}, leading to a
larger tension with data.

\subsection{$\ttbar$ cross-section}\label{sec:xsec}

The good agreement of the inclusive $\ttbar $ cross-section at both
Tevatron and the LHC has been a major constraint on model building for
the $\TAFB$.  Axigluons with purely axial couplings to light and top
 quarks contribute minimally to the total
$\ttbar$ cross-section, but in the flavor-nonuniversal models we
consider, at least one species of quark has non-vanishing vector
couplings to the axigluon. The cross-section constraints are
consequently tighter in these flavor-nonuniversal models.

In Fig.~\ref{5 percent contours} we show contours corresponding to a
5\% and 10\% increase in the LO top pair production cross-section at
the Tevatron for various choices of $g^D_R$ and $g^t_R$, in the
down-type non-universal and top non-universal models.  For our
computation to be meaningful, the ratio of the top pair production
cross-section at higher orders to the LO cross section should be
similar in the SM and in the model with a light axigluon.

We choose $5\%$ as a benchmark because it is comparable to the
combined error on the measurement, which is in agreement with the SM
expectation.  Note that the measured central values are above the predicted NNLO value for $m_t = 173$ GeV \cite{CDFxsec, Langenfeld:2009wd}, so a
$5\%$ increase in the LO Tevatron top pair production cross-section is
perfectly acceptable. 

We superimpose on this figure the CDF $1 \sigma$ preferred regions for the $\TAFB$ (using only the inclusive unfolded measurement).   As mass increases, the global maximum $\TAFB$ decreases, leading to the sharp upward turn of the curves around $2 m_t$.  While the allowed contours around $2 m_t$ appear open at larger couplings, eventually they will close (off the range of the plot), where $\TAFB$ falls back below the measured value $-$ $1 \sigma$ at sufficiently large coupling.  Couplings large enough to provide a good fit to lower $\TAC$ for the RH down-type non-universal model are marginally in agreement with the
$t \bar{t}$ cross-section (see Fig.~\ref{ud nonuniversal scan}). RH
top non-universal models are marginal only in the high mass
range. 

\begin{figure}
\begin{centering}
\begin{tabular}{c}
\includegraphics[width=0.48 \textwidth]{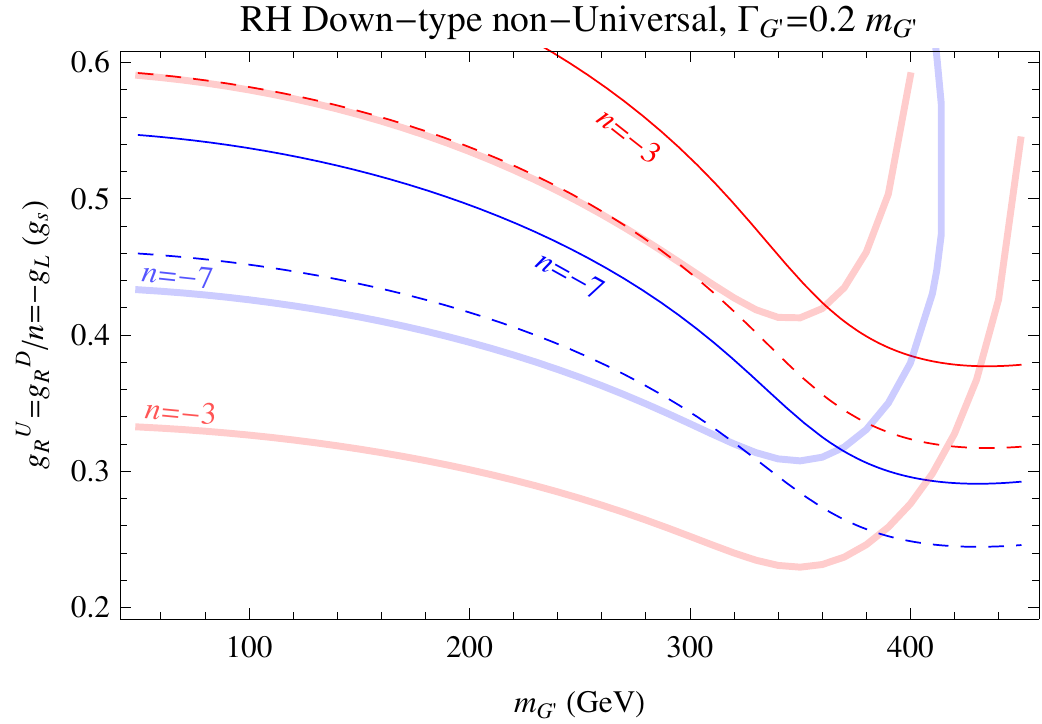} 
\includegraphics[width=0.48 \textwidth]{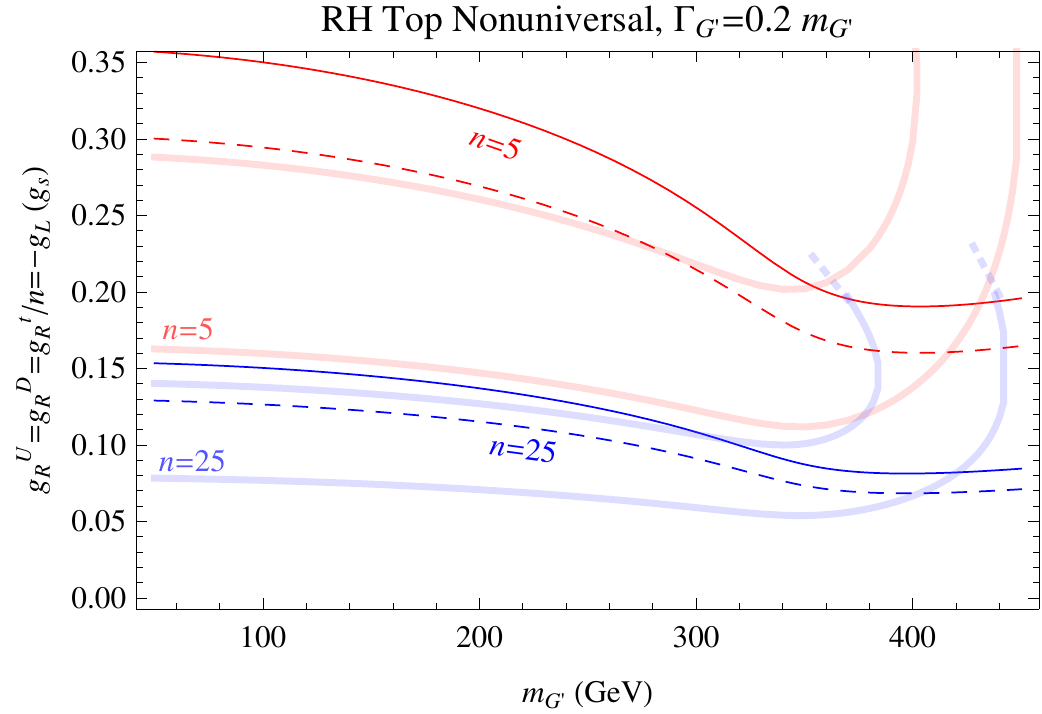} 
\end{tabular}
\end{centering}
\caption{ Contours corresponding to a 5\% (dashed) and 10\% (solid)
  increase in LO top pair production cross-section at the
  Tevatron. For reference, regions between the thick transparent lines
  of the same color indicate the CDF $\TAFB$ 1$\sigma$ preferred
  regions. The $n=25$ region boundaries are drawn precisely only up to
  the strong coupling regime $g_R^t = 25 \times 0.2 g_s$. Note that
  couplings large enough to provide a good fit to lower $\TAC$ for the
  RH down-type non-universal model are marginally in agreement with
  the $t \bar{t}$ cross-section (See Fig.~\ref{ud nonuniversal
    scan}). RH top non-universal models have trouble only in the high
  mass range. }\label{5 percent
  contours}
\end{figure}

\section {Direct searches at hadron colliders}\label{sec:colliders}

Axigluons in the mass range of interest are light enough to have been
copiously produced at past colliders. While in principle
electron-positron colliders and electron-proton colliders can
constrain axigluons, in practice the only existing constraints come
from searches done at hadron colliders. In this section we discuss the
most relevant constraints on axigluons, both broad and narrow, from
various searches done at the SppS, the Tevatron, and the LHC.

\subsection{Narrow Resonances}\label{sec:pprod}

Dijet resonances are constrained by experiments UA1
\cite{Albajar:1988rs} and UA2 \cite{Alitti:1993pn}, dating from the
time of the discovery of the $W$ and $Z$ bosons. At the SppS and also
at the Tevatron, the relevant searches are those looking for single
resonant production of a new state.  For axigluons, single resonant
production occurs through the coupling to quarks\footnote{The
  coupling $g$-$g$-$G'$ occurs only at dimension-six and is
  model-dependent.}, and therefore the production cross-sections
depend in detail on the flavor structure of the model. 

\begin{figure}
\begin{center}
\includegraphics[width=0.65 \textwidth]{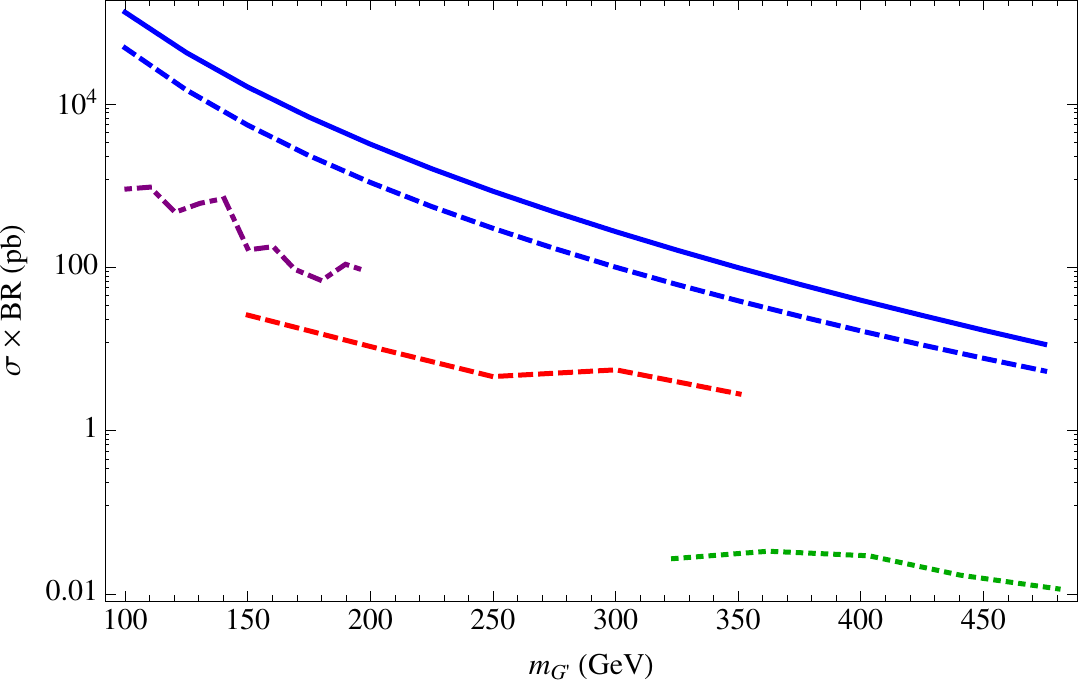} \hspace{2mm} 
\caption{ Rescaled ATLAS 95\% CL and CMS 95\% CL limits on the pair
  production of narrow axigluons.  The red dashed line shows the ATLAS
  5 fb$ ^ {-1} $ limit \cite{ATLAS-CONF-2012-110}, the purple
  dash-dotted line the ATLAS 34 pb$^{-1}$ limit \cite{Aad:2011yh}, and
  the green dotted line the CMS 2.2 fb$^{-1}$ limit
  \cite{CMS-PAS-EXO-11-016}.  The leading order inclusive
  cross-section is shown by the blue solid line.  To highlight the
  model independence of the exclusion we have also shown the LO pair
  production cross-section for $\chi = 0$ in the blue dashed line.
  \label{fig:atlaspprod}}
\end{center}
\end{figure}

Recent LHC searches by both ATLAS \cite{Aad:2011yh,
  ATLAS-CONF-2012-110} and CMS \cite{CMS-PAS-EXO-11-016} have looked
for {\it pairs} of dijet resonances, which probe the QCD pair
production of axigluons through their irreducible couplings to gluons,
$gg\to G' G'$.  Unitarity of the UV completion does not allow
substantial suppression of the axigluon pair production cross-section
below QCD strength: the tree-level non-covariant derivative coupling
of the axigluon to gluons, $\chi$, is fixed by unitarity to the value
$\chi=1$, and even large, order-one loop corrections to $\chi$ are not
sufficient to reduce the pair production cross-section below
experimental bounds.  These searches exclude narrow axigluons in the
entire mass range of interest, independent of the axigluon couplings
to quarks.  ATLAS searches exclude octet (pseudo-)scalars $\phi$ with
masses in the range $100\gev < m_{\phi} < 287\gev$.  To understand how
this limits axigluon pair production, it is necessary to translate the
ATLAS limits on (pseudo-)scalars to (pseudo-)vectors.  This is not
entirely trivial, as the ATLAS searches use control regions to derive
predictions for the background in the signal region, and scalars and
vectors populate the signal and control regions differently.
Fortunately for our purposes, vectors contribute proportionally much
more to the background regions than the scalars do, and thus the
limits on the cross-section derived for the scalar case are
conservative when applied to vectors.  We translate the ATLAS limits
by taking into account the different efficiencies for scalars and
vectors to pass the selection cuts, and show the resulting limits in
Fig.~\ref{fig:atlaspprod}.  Pair production of narrow axigluons is
comfortably excluded over the entire mass range considered by ATLAS,
$100\gev < m_{G'} < 350\gev$.  The exclusion is more stringent than
the exclusion for scalars due to the significantly larger
cross-sections for vectors \cite{Dobrescu:2007yp}.   Note also that
narrow axigluons cannot be ``hidden'' from the search and still remain 
narrow:  suppressing the branching fraction to dijets by the
$\mc{O}(0.1)$ factor necessary to satisfy the exclusions would then require
axigluons to have a total width  $\Gamma_{G'} >0.1\, m_{G'}$.
Meanwhile, the search by CMS excludes octet vectors in the range $320
\gev < m_{G'} < 580\gev$.  Thus, the combination of ATLAS and CMS
searches exclude narrow axigluons above 100 GeV in the entire mass
range under consideration.  

Very recently, a similar search for
axigluon pair production at CDF, $q\bar q \to G' G' \to 4 j$, has
excluded the extremely low-mass region $50\gev < m_{G'} < 125$ GeV, in
the limit of vanishing quark coupling to axigluons \cite{cdf4j}.
While application of this limit to axigluons  which have the quark
couplings necessary to explain $\TAFB$ requires a careful treatment of
quark-initiated contributions to the cross-section, the lack of any
observed excess disfavors such axigluons below 100 GeV.  Such extremely light,
narrow axigluons can also be constrained by bounds on same-sign top production
 from the LHC \cite{AguilarSaavedra:2011ck}.

\subsection{Broad resonances}\label{sec:broadreso}

\begin{figure}
\begin{center}
\includegraphics[width=0.65 \textwidth]{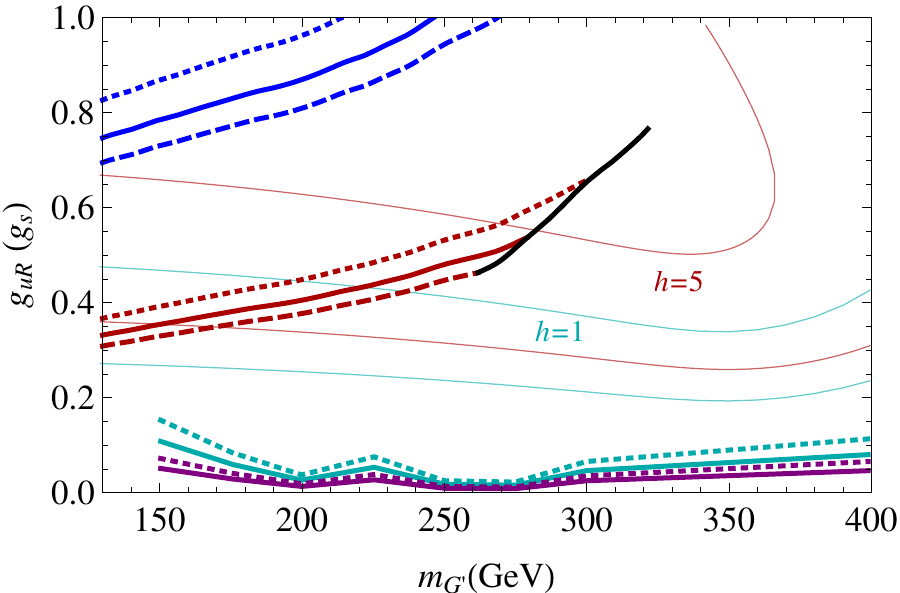} \hspace{2mm} 
\caption{  Topology-dependent CDF and UA1 95\% CL limits on broad axigluons. CDF
limits on $G'\to 2 X \to 4 j$ are shown in cyan for flavor-universal
axigluons and in purple for the scenario where $g_{u_R}=-g_{Q_L}
=-g_{d_R}/5 $, assuming in both cases $BR(G'\to 4 j) = 1$.   Solid lines
 assume that signal
efficiencies are unchanged from the reference axigluon model in
Ref.~\cite{cdf4j}; dotted lines incorporate a $50\%$ reduction in efficiency.
  The blue and red
families of lines denote UA1 limits on $\sigma \times BR(G'\to j j)$
for the flavor-universal case (blue), and the scenario where
$g_{u_R}=-g_{Q_L} =-g_{d_R}/5 $ (red).  Here we have taken into
account that in general $BR(G'\to j j)< 1$ in order to obtain a
sufficiently large total width.  The solid line applies to axigluons
with a fixed total width $\Gamma_{tot} = 0.2 m_G$, the dotted line to
a fixed total width $\Gamma_{tot} = 0.3 m_G$, and the dashed line to a
fixed total width $\Gamma_{tot} = 0.15 m_G$.  In black is the limit on
the down-type non-universal axigluon with $BR(G'\to j j)= 1$, when the
natural width lies in the range $0.15 m < \Gamma_G < 0.4 m$.  Regions
above the lines are excluded.  CDF one-sigma $\TAFB$ preferred regions
are shown for comparison.
\label{fig:dijets}}
\end{center}
\end{figure}

UA1 is the only collaboration to have used dijet searches to set
limits on broad as well as narrow axigluons \cite{Albajar:1988rs},
conducting a dijet search for axigluons with width up to
$\Gamma_{G'}\lsim 0.4 m_{ G'} $.  This search covers the mass range
above $m_{G'} = 150$ GeV, and excludes $g_s$-coupled axigluons up to
$310$ GeV. Rescaling their limits, we obtain the constraints shown in
Fig.~\ref{fig:dijets}, for both flavor-universal and down-type
non-universal scenarios. We use Madgraph to obtain the relative
fraction of down- and up-initiated events.  Note that, in the majority
of parameter space, the natural width into dijets is not sufficient to
make the axigluon broad ($\Gamma \gsim 0.15 m_G$), and in rescaling
the limits we must therefore allow for non-zero branching fractions
into undetected final states.

As a caveat, note that the UA1 study modeled the longitudinal and
transverse momentum distributions of the $G'$ using a sequential $Z'$.
The slight difference between the $G'$ and $Z'$ in the up versus down
PDF support of the inclusive production cross-section does lead to a
slight (percent level) change in efficiencies due to the different
rapidity distributions of the center of mass.  Of more concern is the
difference in the transverse and longitudinal momentum distributions
due to the different ISR spectra of a color octet versus a color
singlet. However, as the cross-section UA1 used to set limits is
leading order, the limits should be reliable.

Increasing the down-type coupling to alleviate tension with the CMS
$\TAC$ measurement tightens dijet constraints significantly, while a
moderate increase in the top-quark coupling could allow for acceptably
large values of $\TAFB$ with couplings to light quarks small enough to
evade dijet constraints.

The CDF paired dijet search \cite{cdf4j} can also constrain broad
axigluons if they decay dominantly according to the cascade $G'\to X X
\to 4 j$, and if the axigluon width is not substantially larger than
the experimental resolution.\footnote{Ref.~\cite{Gross:2012bz} has
  made a similar argument regarding the LHC paired dijet searches.}
We show limits from the CDF exclusion in Fig.~\ref{fig:dijets}.  We
have used the data for the case where the intermediate $X$ has mass
$m_X = 50$ GeV, but for fixed $m_{G'}$ the cross-section limits do not
depend strongly on $m_X$.
In this search experimental resolutions on the four-jet invariant mass
are on the order of $25\%$; larger axigluon widths will make it more
difficult to obtain an accurate background estimate and fit a
localized signal template.  However, it can be seen from
Fig.~\ref{fig:dijets} that even after reducing the signal efficiency
by $50\%$, axigluons decaying dominantly into dijets are eliminated as
a possible explanation for $\TAFB$.

\subsection{ Constraints on daughter particles}

To evade the LHC pair production exclusions, any light axigluon
explanation for $\TAFB$ must necessarily be either less than 100 GeV
in mass \cite{Krnjaic:2011ub} or sufficiently broad, with sufficiently
small branching fraction into dijets, to fail the selection cuts
\cite{Tavares:2011zg}.  In the window below 100 GeV, the tensions with
PEW constraints we consider in the next section are important.  Thus
broad axigluons are the only states remaining that are obviously
consistent with the data.  In many regions of parameter space,
however, the axigluon-SM couplings do not yield a large axigluon width
($\Gamma\gsim 0.1 m_{G'}$), necessitating the introduction of new
colored degrees of freedom to provide a BSM decay mode for the
axigluon. The nature of these new degrees of freedom is highly model
dependent, but in many cases they may be easier to search for than the
axigluon itself.  For example, the paired dijet searches discussed in
\ref{sec:pprod} exclude the possibility of axigluon decay into pairs
of octet scalars for axigluons in the mass range $200 \gev < m_{G'} <
574 \gev$. CDF \cite{cdf4j} excludes triplet scalars decaying to
dijets in the mass range between 50 GeV and 100 GeV. In addition to
the paired dijet searches, both CDF \cite{Aaltonen:2011sg} and CMS
\cite{Chatrchyan:2011cj,:2012gw} have conducted searches for three jet
resonances, excluding octet fermions in the mass ranges from $70\gev <
m_f < 145\gev$ and $ 200\gev < m < 460\gev$ respectively, thereby
constraining the decays of axigluons involving three-jet
resonances. Other possibilities, involving longer or less symmetric
decay chains, are less constrained.  A detailed discussion of decay
scenarios in light axigluon models and relevant constraints is
provided in \cite{Gross:2012bz}.

\section{Precision electroweak}\label{sec: pew}

The strongest precision electroweak constraints on $G' q \bar{q}$
couplings arise from the one-loop corrections to the $Z q \bar{q}$
vertex.  These corrections act uniformly to increase the effective $Z
q \bar{q}$ couplings, leading to significant constraints from the
hadronic $Z$ width and the hadronic $Z$ pole production cross-section,
$\sigma_\text{had}$.  The related real emission process, $Z
\rightarrow G' q \bar{q}$, is relevant when $m_{G'} < m_Z$ and should
also be taken into account.  Contributions from $Z b \bar{b}$
observables and the $S$ and $T$ parameters are subdominant, as they
are for heavy axigluons \cite{Haisch:2011up}.

We recomputed the one-loop corrections to the $Zq\bar q$ vertices and
incorporated corrections for large axigluon widths.  Broad widths tend
to \emph{increase} the contribution to the hadronic $Z$ width,
particularly close to the $Z$ mass and below.  In this region, a large
axigluon width increases the contribution to the hadronic $Z$ width by
(for example) about 5\% at $m_Z$ for a 40\% width.  Much above the $Z$
mass, broad widths minimally affect PEW corrections.  The contribution
to the hadronic $Z$ width from real emission of axigluons below the
$Z$ mass can be approximated from the expression in
\cite{Carone:1994aa}.  Further details on the inclusion of the width
and the extraction of the real emission contribution can be found in
Appendix \ref{sec: axiglue correction}.

\begin{figure}
\includegraphics[width=0.65 \textwidth]{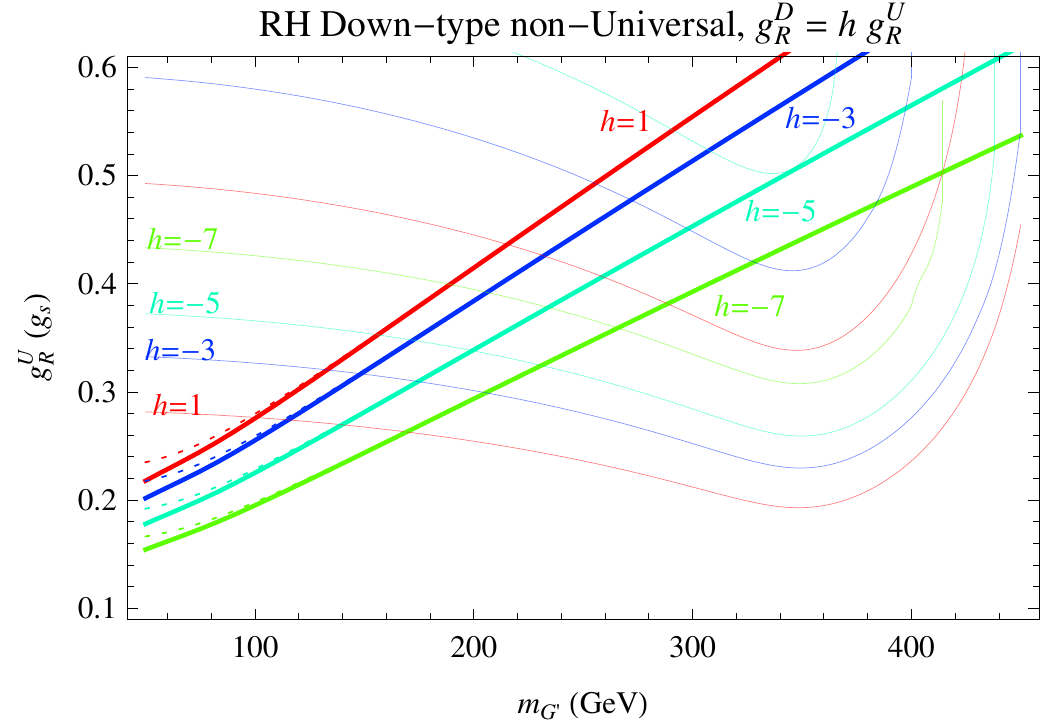}
\caption{ 95\% C.L.  exclusion contours from $\sigma_\text{had}$ and
  $\Gamma_Z$ measurements at LEP for axigluons with purely axial
  couplings to up-type quarks and modified couplings to down-type
  quarks of the form $g_R^D = h g_R^U$. Solid curves correspond to
  including a width $\Gamma_{G'} / m_{G'} = 0.2$ (see Appendix
  \ref{sec: axiglue correction}) and dotted curves (visible only at low mass) correspond to the
  zero width limit. As can be seen, the effect of the finite width is
  very
  small.
  An additional contribution from axigluon emission for masses below
  $m_Z$ is included; see the text for a discussion.  For reference,
  the corresponding boundaries of the CDF $\TAFB$ 1$\sigma$ preferred
  region (thin, curves of corresponding color) are shown.
}\label{pew}
\end{figure}

As in \cite{Haisch:2011up}, to derive constraints we use the combined
LEP results on $\Gamma_Z$ and $\sigma_\text{had}$ assuming lepton
universality \cite{ALEPH:2005ab}. We use the same SM inputs as
\cite{Haisch:2011up}.  The resulting $95\%$ C.L.~exclusions for 0\%
and $20\%$ widths are plotted in Fig.~\ref{pew}.  We show contours for
couplings of the form $g_R^D = h g_R^U$. The $h=1$ contour corresponds
to the flavor universal case as well as the top non-universal case
since the top coupling does not enter the correction. Limits for
masses below $m_Z$ should be considered cautiously in light of the
fact that in this mass range the axigluon can affect the running of
$\alpha_s$ and thus also the extraction of SM parameters used in the
calculation of $\Gamma_\text{had}$. On the other hand, careful
analysis of the running of $\alpha_s$ would provide yet another bound
in this mass range.  To estimate possible ambiguities associated with
extracting $\alpha_s$ we include a curve assuming 1 \% decrease in
$\alpha_s$ for $m_{G'} < m_Z$ in the summary plots of \S\ref{sec:
  summary} (see Table \ref{tab: constraints} and Figs.~\ref{fig:
  summary curves: flavor universal}-\ref{fig: summary curves: flavor
  nonuniversal}).

The universal axigluon (top red line in Fig.~\ref{pew}) with couplings
sufficient to reproduce the measured $\TAFB$ is excluded below 100
GeV. PEW constraints rule out larger regions of parameter space as $h$
becomes more negative. For instance, models with $g^D_R \lesssim -5
g^U_R$ appear to be in tension with PEW measurements for masses below
about 200 GeV.

\begin{figure}
\includegraphics[height=0.15\textheight]{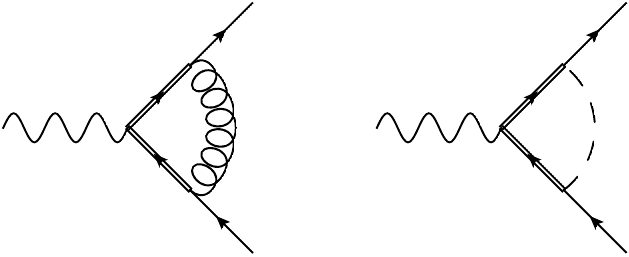}
\caption{Representative diagrams involving heavy vector-like quarks $Q_h$
  (double lines) and scalar $\hat \phi$ (dashed lines) which contribute to the $Z q \bar{q}$ vertex
  correction.}\label{fig: Zqq diagrams}
\end{figure}

As the PEW constraints are due to loop corrections, they can change
depending on additional UV content in the model.  It is therefore of
interest to ask how the PEW constraints depend on the minimal UV
completion of the phenomenological axigluon model.  A UV-complete
description of an axigluon with small ($< g_s$) axial couplings to
quarks necessarily requires new heavy fermion degrees of freedom, as
reviewed in Appendix \ref{sec:UV}.  Loops of heavy fermions, $Q_h$,
and the axigluon can contribute to the vertex correction, as in
Fig.~\ref{fig: Zqq diagrams}.  Once the axigluon mass and light quark
couplings are fixed, the free parameters are the masses of the heavy
fermions and the gauge mixing angle $\tan\theta$; the heavy fermion
couplings to the axigluon and to the light quarks are otherwise
determined.  The new fermions, in particular, have been proposed as
possible new decay modes for the axigluon
\cite{Tavares:2011zg,Gross:2012bz}, requiring at least one flavor to
be light, and thus relevant for the PEW calculation.

We have computed the contribution from heavy fermions to the $Zq\bar
q$ vertex correction and find that the sign of the contribution is the
same as that of the correction from axigluon and light quarks alone.
The PEW bounds from an axigluon alone are thus \emph{conservative}.
PEW bounds for a few representative choices of heavy quark mass (and
flavor) are shown in Fig.~\ref{fig: heavyq}.  More details of the calculation can be
found in the appendix \S \ref{sec: heavy quark correction}.

\begin{figure}
\includegraphics[width=0.49\textwidth]{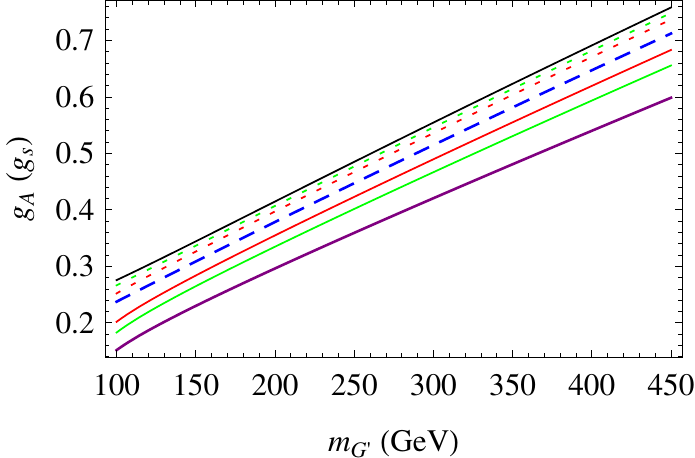}
\includegraphics[width=0.49\textwidth]{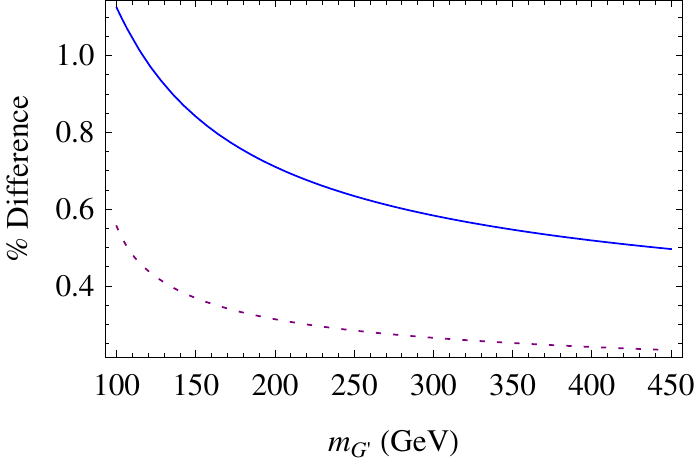}
\caption{\emph{Left:} 95\% C.L.  exclusion contours from
  $\sigma_\text{had}$ and $\Gamma_Z$ measurements at LEP for
  flavor-universal axigluons in the presence of no species (solid
  black, top), one species of heavy quark from the UV completion with 
  coupling to $u_R$ (red dotted), $u_L$ (solid red), $d_R$ (green
  dotted---essentially on top of no species curve), $d_L$ (solid
  green), and to all four first-generation species $u_R, u_L, d_R,
  d_L$ (purple). For these curves we take the new fermions to have mass $m_{Q_h} =
   m_{G'}/2$. With this choice the contribution of the scalar
  is imperceptible on the plot, at least for scalar masses
  $m_{\hat{\phi}} \geq m_{Q_h}$; for smaller masses real emission
  would come into play.  The blue dashed curve shows the exclusion contour  given $m_{Q_h} =
   4 m_{G'}$ and heavy quarks coupling to $u_R, u_L, d_R,
  d_L$, not including the scalar contribution.   We have taken the coupling strength of the new
  fermion to standard model quark and axigluon to be double the
  axigluon-quark-quark coupling: $g_\text{mix} = 2 g_A$.  See the
  appendix \S\ref{sec: heavy quark correction} for
  details. \emph{Right:} Percent difference between constraint curves $g_A (m_{G'})$
  including and not including scalar contributions, in the presence of
  four heavy quarks from the UV completion coupling to $u_R, u_L, d_R,
  d_L$, with $m_{Q_h} = m_{\hat{\phi}} = m_{G'}/2$ (blue), and with $m_{Q_h} = 4 m_{G'}$, $m_{\hat{\phi}} = m_{G'}/2$ (dotted purple).
  Here again we choose $g_\text{mix} = 2 g_A$ for purposes of
  illustration. }\label{fig: heavyq}
\end{figure}

Typical UV completions also contain a neutral scalar $\hat\phi$, the
uneaten remnant of the field responsible for spontaneous symmetry
breaking.  Like the axigluon, this scalar also has off-diagonal
$Q_h$-$q$ couplings with a fixed strength, and can contribute to PEW
corrections via the right-hand diagram in Fig.~\ref{fig: Zqq
  diagrams}. The calculation of this correction is also presented in
\S \ref{sec: heavy quark correction}. We find that it has the
\emph{opposite} sign as that from the axigluon-light quark loop and so
therefore could serve to moderate precision electroweak
corrections. On the other hand, the coupling entering the correction
is related to that entering the heavy quark correction times the ratio
of new heavy fermion to axigluon mass (see Eq.~\eqref{eq: gmix scalar
  relationship}). If new fermions are light enough to increase the
axigluon widths, the scalar contribution is subdominant. This is shown
in the right-hand panel of Fig.~\ref{fig: heavyq}, where it can be
seen that, while the scalar contribution does weaken the PEW
constraint, it is a {\em very} mild effect in comparison to the
effects of loops of fermions.

\section{Results and Conclusions}\label{sec: summary}

We have examined constraints on light axigluon models for the Tevatron
top forward-backward asymmetry from the LHC charge asymmetry, dijet
and multijet searches, and precision electroweak observables. We
considered only broad axigluons, as paired dijet resonance searches are
devastating for narrow axigluons, regardless of the flavor structure.

\begin{table}
\begin{tabular}{r @{: \;} p{2.8 in} @{~~} p{2 in}}
Name & Curve(s) plotted & Value fitted   \\
\hline
CDF AFB &  $\pm 1 \sigma$ band (solid dark blue),  \cite{Aaltonen:2012it} & $0.164 - 0.066 \pm 0.045 \pm 0.020$\\
D0 AFB &  $\pm 1 \sigma$ band (solid cyan), \cite{Abazov:2011rq} & $0.196-0.063\pm0.065\pm0.019$  \\
CMS AC & $+1 \sigma$ curve (solid purple), preferred region below, \cite{CMS-PAS-TOP-11-030} & $0.004-0.0115\pm0.0156\pm0.0006$ \\
ATLAS AC & $\pm 1 \sigma$ band (solid green), \cite{ATLAS-CONF-2012-057} & $0.029-0.006\pm0.023\pm0.002 $\\
PEW & LEP precision electroweak using $\sigma_\text{had}$ and $\gamma_Z$, 95\% C.L. exclusion (solid black, dashed black for ${\delta \alpha_s \over \alpha_s}\sim-1\%$ below $m_Z$),  \cite{ALEPH:2005ab,Haisch:2011up} &  $\Gamma_Z^\text{exp.} = 2.4952 \pm 0.0023$ GeV, \; \; \;  $\sigma_\text{had}^\text{exp.} = 41.540\pm0.037$ nb,  $\text{correlation}=\left(\begin{smallmatrix} 1 & -0.3 \\ -0.3 & 1 \end{smallmatrix} \right)$,  $\Gamma_Z^\text{SM} = 2.4945 \pm 0.0007$ GeV,  \; \; \; $\sigma_\text{had}^\text{SM} = 41.482\pm0.006$ nb\\
Tevatron $\sigma$ & 10\% over LO SM $\sigma_{t \bar{t}}$ (reddish-pink, solid), off scale on most plots & $5.7 \text{pb} \pm 10\%$ \\
CDF high-mass AFB &  $\pm 1 \sigma$ band  (dashed dark blue), \cite{Aaltonen:2012it} & $0.295-0.1\pm0.066\pm0.03 $\\
UA1 dijets (broad) & 95\% exclusion (solid brown), \cite{Albajar:1988rs}\\
\hline
\end{tabular}
\caption{Curves plotted in Figs.~\ref{fig: summary curves: flavor universal} 
  and \ref{fig: summary curves: flavor nonuniversal}. References for experimental 
  inputs are noted where relevant. For the charge asymmetry measurements ``1$\sigma$''
  is taken to be the combined statistical $\oplus$ systematic experimental $\oplus$ 
  Standard Model expectation error. Unless noted otherwise, the ``Value fitted'' 
  is (experiment central value)$-$(SM prediction)$\pm$(experimental error)$\pm$(SM 
  prediction error). For each curve, we take the Standard Model prediction and
  associated error as quoted by the collaboration, \emph{except} we apply a
  correction for electroweak contributions to D0's values. Following CDF, we 
  multiply the SM NLO expectation by 1.26 and estimate a 30\% error on the 
  expectation.}\label{tab: constraints}
\end{table}

Besides the Tevatron measurements, the most important constraints come
from the LHC charge asymmetry and precision electroweak observables.
The LHC charge asymmetry has the potential to severely constrain light
axiglue models for $\TAFB$, but the current spread in the central
values measured by ATLAS and CMS leaves the situation unsettled.
Future evolution towards the small values preferred by the current
semileptonic measurement of CMS would be devastating for flavor-universal
models.  Precision electroweak constraints eliminate a significant
corner of the very low-mass parameter space.

\begin{figure}
\centering
\includegraphics[width=0.83 \textwidth]{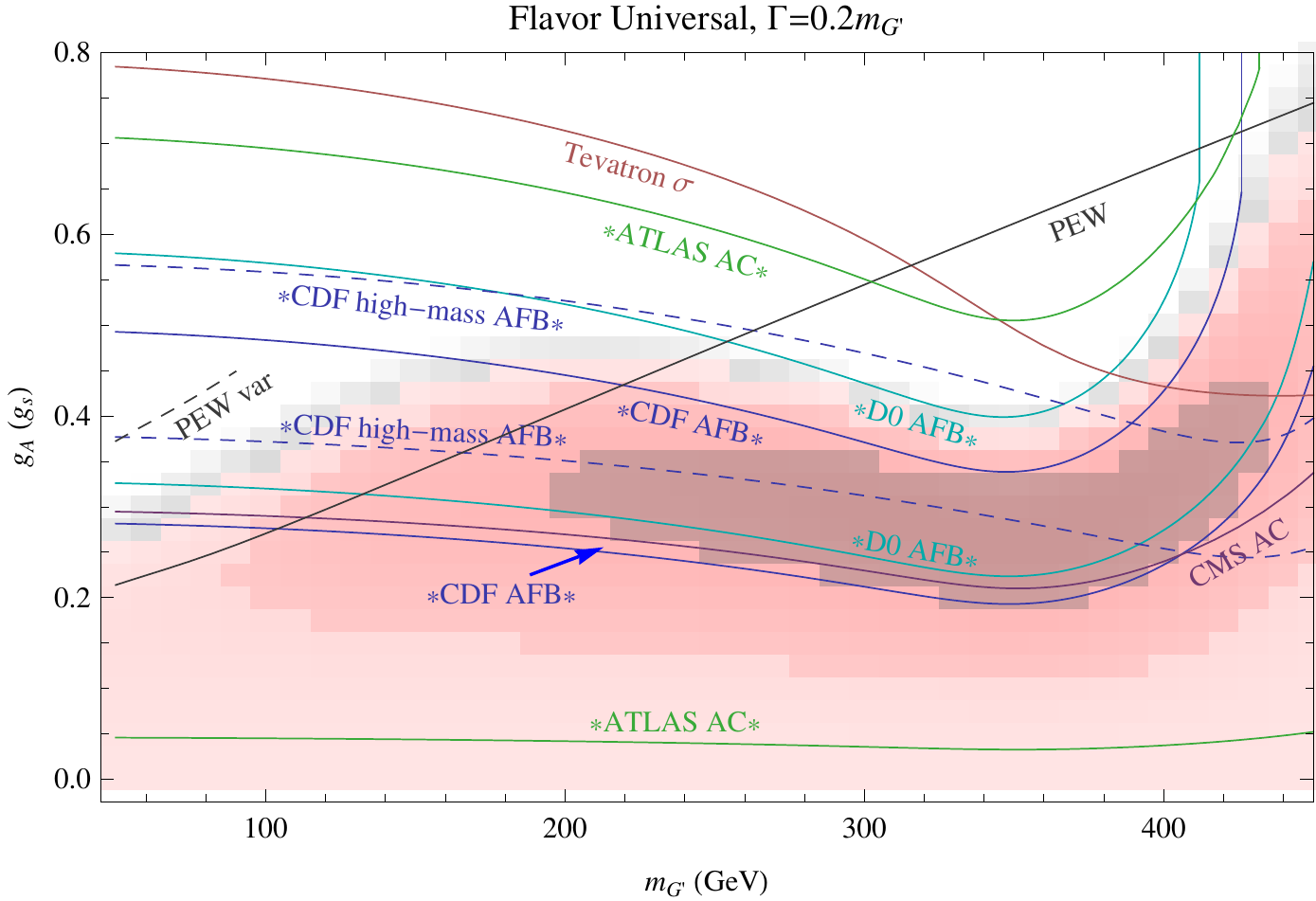} \\
\includegraphics[width=0.7\textwidth]{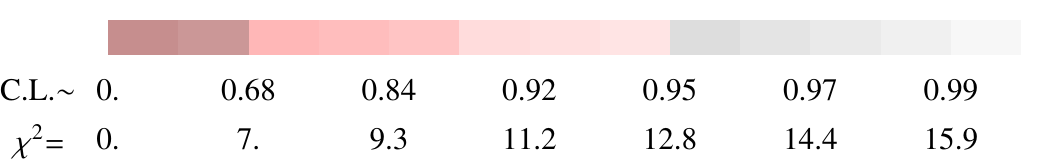}\\
\caption{Summary of constraints on flavor-universal axigluons in the
  mass-coupling plane. Contours correspond to those listed in Table
  \ref{tab: constraints}. We plot $1 \sigma$ contours for asymmetry
  parameters and $95\%$ exclusion curves for precision electroweak
  (PEW); for the Tevatron cross-section, we use a 10\% increase of LO
  $\sigma_{t \bar{t}}$ as a benchmark. The 95\% C.L.~UA1 broad dijet
  resonance search constraint is off the plot.  Curve labels sit on
  the \emph{preferred} side of the boundary, and curves that are part
  of a band on the plot are indicated by asterisks. Note that for
  axigluon masses in the $2 m_t$ range, top asymmetries can attain a
  global maximum at moderate coupling strengths, which gives rise to a
  sharp upward turn of the asymmetry bands near $2 m_t$. The bands
  close off of the plot. See the discussion in
  \S\ref{sec:xsec}. The $\chi^2$ value
  computed using the first six measurements listed in the right-hand
  column of Table~\ref{tab: constraints} is superimposed. (The
  correlation between $\sigma_Z$ and $\Gamma_Z$ is taken into account
  in the fit.)  } \label{fig: summary curves: flavor universal}
\end{figure}

\begin{figure}
\centering
\includegraphics[width=0.85 \textwidth]{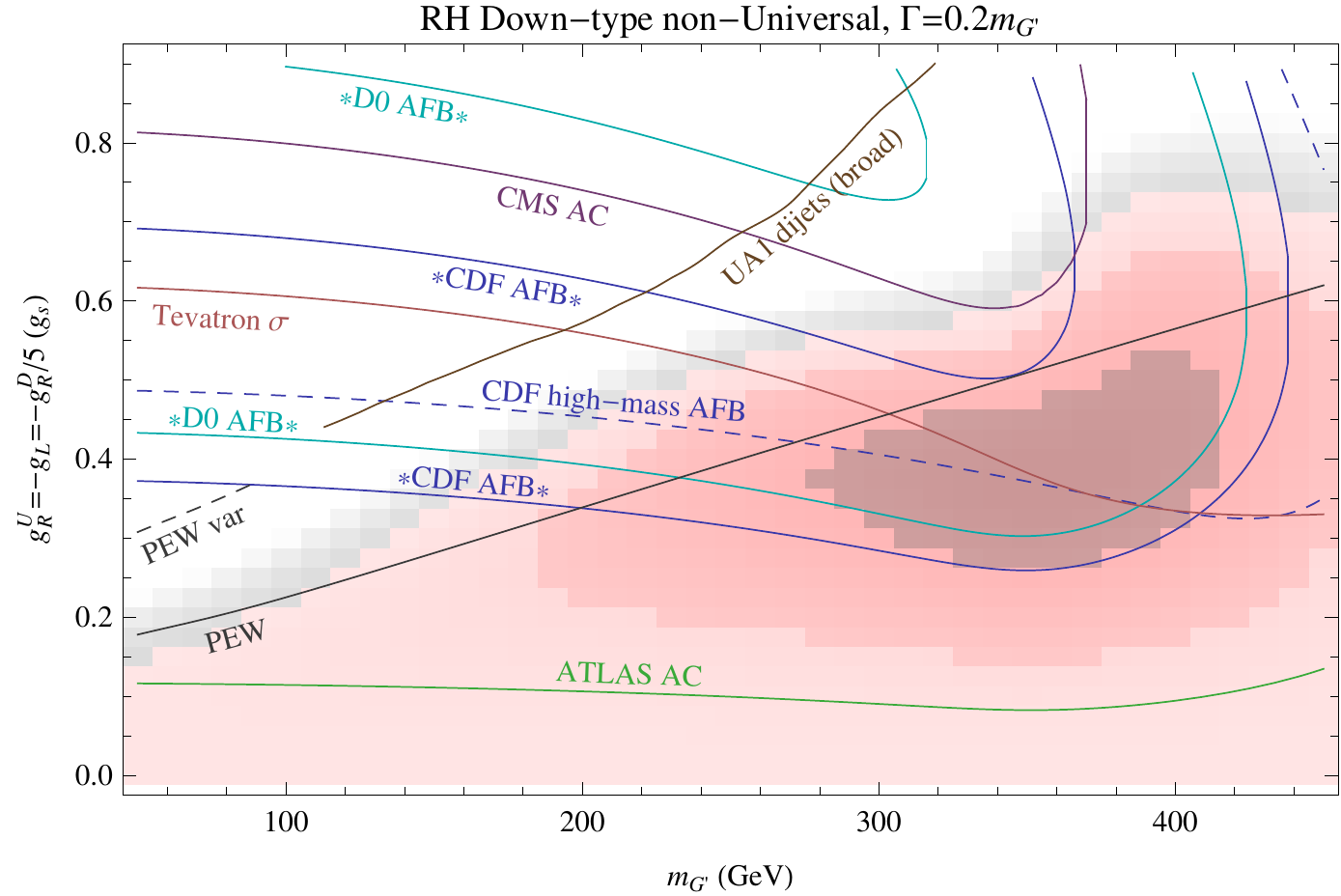} \\ 

\vspace{0.5 cm}

\includegraphics[width=0.85 \textwidth]{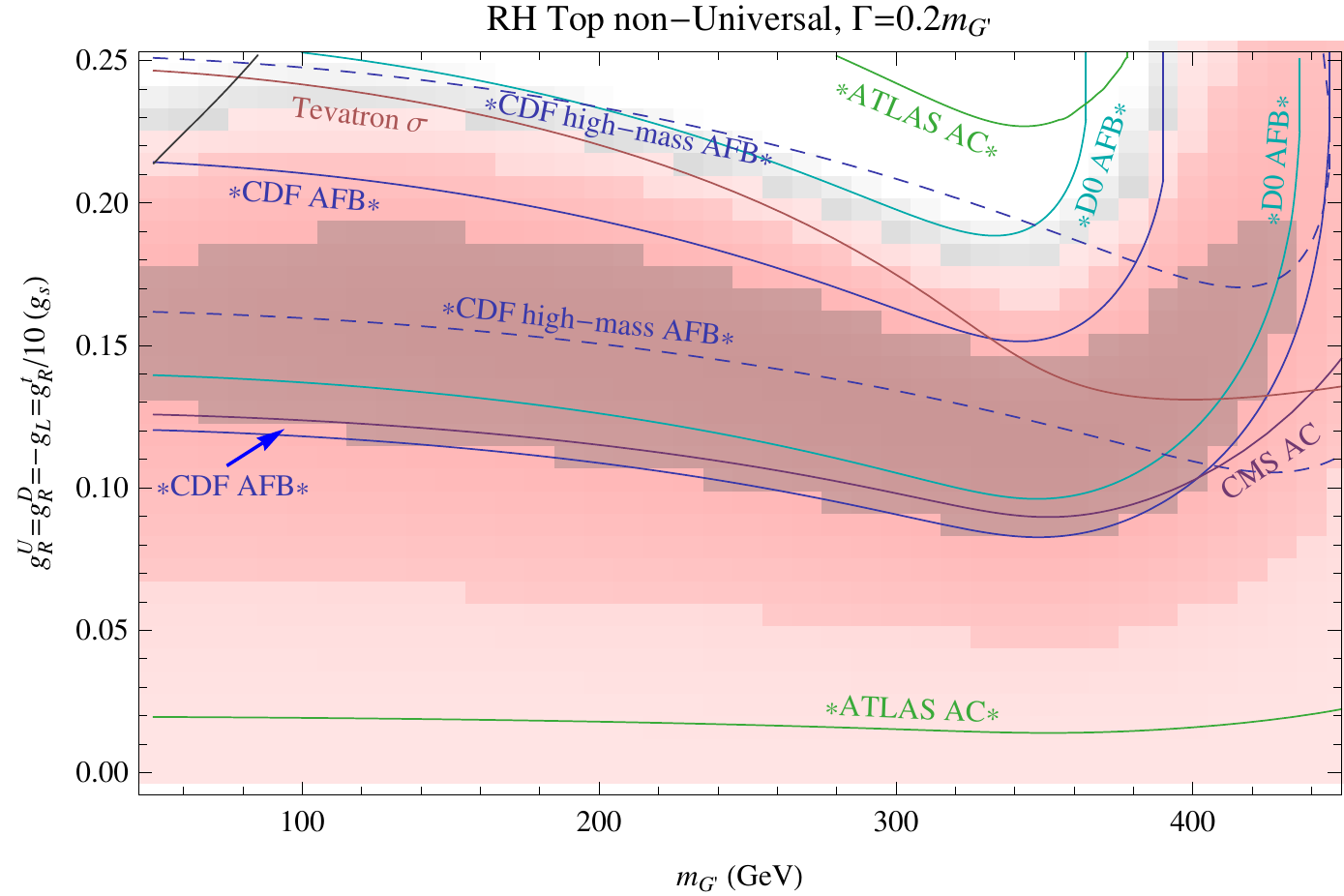} 
\caption{Summary of constraints for RH down-type non-universal
  axigluons with $g_R^D = -5 g_R^U$ (top) and RH top non-universal
  axigluons with $g_R^t = 10 g_R^U$ (bottom). Contours correspond to
  those listed in Table \ref{tab: constraints}; refer also to
  Fig.~\ref{fig: summary curves: flavor universal}. Note that the RH
  down-type non-univeral model can simultaneously satisfy the CMS
  $\TAC$ constraint and Tevatron $\TAFB$ constraints (see also
  Fig.~\ref{ud nonuniversal scan}) at $1 \sigma$, in contrast to the
  flavor-universal and RH top non-universal
  models. 
}\label{fig: summary curves: flavor nonuniversal}
\end{figure}

We considered, in addition to flavor-universal axigluons, two flavor
structures that can ameliorate either one of these constraints.
Down-type nonuniversal models can ameliorate tension between LHC and
Tevatron top asymmetries, but are significantly more constrained by
PEW; top non-universal models can evade constraints from PEW, but do
not help with the tension with the LHC $\TAC$.  In addition, top
non-universal models are constrained by measurements of the lepton
asymmetry at the Tevatron.

Our conclusions are shown in the plots Figs.~\ref{fig: summary curves:
  flavor universal} and \ref{fig: summary curves: flavor
  nonuniversal}, which show the parameter space consistent with all
constraints at 1$\sigma$. The PEW constraints shown are $95\%$
C.L. and the Tevatron cross-section curve corresponds to a $10\%$
increase above the Standard Model in the leading order
cross-section. The contours are superimposed on a granular density
plot of a $\chi^2$ fit using the CDF and D0 measurements of the
inclusive $\TAFB$, the ATLAS and CMS $\TAC$ measurements, and LEP's
combined measurement of $\sigma_\text{had}$ and $\Gamma_Z$. In all we
used 6 inputs, with the only cross-correlation being between
$\Gamma_Z$ and $\sigma_\text{had}$. 

Note in particular that for the flavor-universal and top nonuniversal
models, the globally preferred region lies outside the 1-sigma band
preferred by the CMS $\TAC$ measurement almost everywhere. This
highlights the potential power of the LHC charge asymmetry
measurements. Out of all the indirect constraints considered here, a
reduction of a factor 2 in the error bars of the LHC charge asymmetry
measurement will have the most impact in ruling out the remaining
regions of parameter space.   The down-type nonuniversal models which
 can be brought into agreement with the CMS asymmetry measurement
 encounter instead accentuated difficulty with PEW constraints.

Axigluons remain one of the best options for explaining the Tevatron
forward-backward asymmetry with new physics. Much of the parameter
space has been closed, in particular for narrow axigluons, and
additional avenues should also be sought to explain the signal
observed by Tevatron. While direct searches for broad axigluons are
challenging and model dependent, we have shown that indirect
observables are capable of tightly constraining admissible windows
without reference to any specific decay scenario.  Moreover, the
combination of constraints from LHC charge asymmetries, PEW
observables, and the lepton asymmetry leaves no obvious flavor avenue
open to escape the tightening net of indirect constraints.

\bigskip

{\em Acknowledgements:} It is a pleasure to thank B.~Dobrescu,
M.~Schmaltz, S.~Westhoff, and D.~Whiteson for useful conversations. JS
thanks A.~Falkowski and M.~Schmaltz for collaboration on a related
project. JS was supported by DOE grant DE-FG02-92ER40704, NSF grant
PHY-1067976, and the LHC Theory Initiative under grant
NSF-PHY-0969510. KZ is supported by NSF CAREER award PHY 1049896 as
well by the DOE under grant de-sc0007859.  JS and KZ thank the Aspen
Center for Physics under grant NSF-PHY-1066293, as well as the Galileo
Galilei Institute and the INFN, for hospitality and partial support
during the completion of this work.

\appendix

\section{Minimal UV Completions}\label{sec:UV}

In this section we review how to obtain axigluon models with small
quark-axigluon couplings from a UV-complete description of spontaneous
symmetry breakdown.  We will neglect considerations of anomaly
cancellation.

As discussed in section~\ref{sec:model}, the minimal symmetry breaking
structure that can realize a massive octet vector is $SU(3)_1\times
SU(3)_2\to SU(3)_c$.  Taking the breaking to be due to the vacuum
expectation value of a bifundamental $\langle\phi\rangle =
f\mathbbm{1}$ and denoting the coupling constants of the two groups as
$g_1<g_2$, the strong coupling constant is, as usual,
\beq
g_s = \frac{g_1g_2}{\sqrt{g_1^2+g_2^2}} \equiv g_1\cos\theta ,
\eeq
 while the axigluon, $G'$, and SM gluon, $g$, are given by the following linear combination of UV gauge fields
\barray
G'^\mu &=& \sin\theta G_1 ^\mu - \cos\theta G_2 ^\mu\\
g^\mu & = &\cos\theta G_1 ^\mu +\sin\theta G_2 ^\mu.
\earray
The gluon remains massless, while the axigluon obtains a mass
\beq
m_{G'} = \sqrt{g_1^2+g_2^2} f .
\eeq
Quark-axigluon couplings depend on the embedding of the SM quarks in
the group $SU(3)_1\times SU(3)_2$.  First consider a (Weyl) quark $Q$
transforming as a fundamental under $SU(3)_1$. After spontaneous symmetry breakdown, its
 coupling to the axigluon is
\beq
\mc{L}_{q1} = g_s \tan\theta \, G'_\mu \, Q^{\dag}\bar\sigma^{\mu} Q .
\eeq
Meanwhile, a quark $Q$ transforming as a fundamental under $SU(3)_2$
couples to the axigluon as
\beq
\mc{L}_{q2} = - g_s \cot\theta \, G'_\mu \,Q^{\dag}\bar\sigma^{\mu} Q .
\eeq
Since if the left-handed fields couple to $SU(3)_1$, the right-handed fields must couple to $SU(3)_2$ (or vice-versa) in order to get axial couplings to $G'$, we can thus see immediately that couplings of the left- and right-handed fields to the axigluon cannot both be smaller than $g_s$. 
It is therefore necessary to introduce heavy fermions that can mix
with the SM quark fields and modify their axigluon couplings
\cite{Dobrescu:2007yp,Tavares:2011zg,Cvetic:2012kv}.

For definiteness consider the case where $Q$ is a fundamental under
$G_1$ and introduce $\hat Q, \bar Q$ transforming as a fundamental and
an anti-fundamental respectively under $G_2$, such that $\hat Q$ has
the same $SU(2)_L\times U(1)_Y$ quantum numbers as $Q$, and $\bar Q$
has the same $SU(2)_L\times U(1)_Y$ quantum numbers as $Q^\dag$.
Then mixing can be obtained through the Lagrangian
\beq
\mc{L}_{mix} = \bar Q \left( M \hat Q + \lambda \phi Q\right) + \hc,
\eeq
where $\phi$ is the field responsible for the spontaneous breakdown of
$SU(3)_1\times SU (3)_2\to SU(3)_c$.  When $\phi$ picks up its vev,
$\phi = f + \sqrt{1/6}\,\hat \phi$, the resulting Lagrangian is
\beq\label{eq:lmix}
\mc{L}_{mix} = \sqrt{M ^ 2+\lambda ^ 2 f ^ 2}\, \bar Q Q_h + \frac{\lambda \cos\alpha}{\sqrt{6}}\,  \hat \phi \bar Q q + \ldots  +\hc,
\eeq
where the mass eigenstates $Q_h$, $q$ are given by
\barray
Q_h &=& \cos\alpha \, \hat Q + \sin\alpha \, Q \\
q & = & - \sin\alpha\,\hat Q +\cos\alpha\, Q
\earray
in terms of the mixing angle
\beq
\cos\alpha =\frac{M}{\sqrt{M ^ 2+\lambda ^ 2 f ^ 2}}.
\eeq
Note that
\beq
m_{Q_h} = \sqrt{M^2 + \lambda^2 f^2}.
\eeq 
The couplings of the different quark states to the two vector states can now be
read off from the kinetic terms,
\barray
\label{eq:fulloffdiag}
\frac{1}{g_s} \mc{L}_{axi} & = &  Q_h ^\dag \bar\sigma^\mu Q_h \left( g_\mu + 
              (-\cos^2\alpha\cot\theta +\sin^ 2\alpha\tan\theta) G_\mu\right) + \\ \nonumber
            &&      q ^\dag \bar\sigma^\mu q\left( g_\mu + 
              (-\sin^2\alpha\cot\theta +\cos^ 2\alpha\tan\theta) G_\mu\right) + \\ \nonumber
            &&   \left( Q_h^\dag\bar\sigma ^\mu q + q^\dag\bar\sigma ^\mu Q_h \right)
                     \left(\cos\alpha\sin\alpha(\cot\theta+\tan\theta) G_\mu \right) .
\earray
The mixing angle $\cos\alpha$ is the necessary ingredient that allows
us to freely dial the quark couplings to axigluons in the
phenomenological low-energy Lagrangian.  However, once $\cos\alpha$
(and $\cot\theta$) are fixed, so are the off-diagonal $q$-$Q$-$G'$
couplings.  This is particularly important for computing precision
electroweak constraints, as we will discuss in the following section.

\section{ Corrections to the $Z$-$q$-$\bar q$ vertex}

\subsection{One-loop corrections with finite axigluon width}\label{sec: axiglue correction}

In unitary gauge a convenient expression for the re-summed axigluon
propagator is \cite{Baur:1995aa}
\begin{align}
D^{\mu \nu}_{G'}(q) &= {-i \over q^2 - m_{G'}^2 + i q^2 \gamma_{G'} } \left( g^{\mu \nu} - {q^\mu q^\nu \over m_{G'}^2} (1 + i \gamma_{G'}) \right) \\
&={1 \over 1 + i \gamma_{G'}} {-i \over q^2 - M_{G'}^2  } \left( g^{\mu \nu} - {q^\mu q^\nu \over M_{G'}^2}  \right) 
\end{align}
where $\gamma_{G'} = \Gamma_{G'} / m_{G'}$ and we have defined 
\beq
M_{G'}^2 \equiv {m_{G'}^2 \over 1 + i \gamma_{G'}}.
\eeq
Since all one-loop precision electroweak corrections of interest
involve exactly one axigluon propagator, the finite width amplitude is
thus related to the one-loop amplitude in the zero-width
approximation, $\mathcal{M}_{0}(m_{G'}^2)$, via
\beq 
\mathcal{M}(m_{G'}^2, \gamma_{G'}) = {1 \over 1 + i \gamma_{G'} } \mathcal{M}_{0}(M_{G'}^2).  
\eeq
Since the Feynman prescription for handling poles is equivalent to
assuming a small positive width, this prescription is consistent.

We calculated the one-loop correction to the $Z q \bar{q}$ vertex and
fermion field strength corrections (in unitary gauge, assuming
massless SM quarks in the loop) and find, in agreement with
\cite{Haisch:2011up}, that in the zero-width limit the correction to
the $Z q_P \bar{q}_P$ coupling, $f_P^q$, is
\beq\label{eq: vertex correction}
 \delta f_P^q =  f_P^q {{g_P^q}^2 \over (4 \pi)^2 } c_F \,K(z)
\eeq
where $z=m_Z^2/ m_{G'}^2$, $c_F = {4 \over 3}$, $g_P^q$ is the axigluon coupling to $q_P
\bar{q}_P$ ($P = R$ or $L$), and $K (z)$ is 
given by
\barray
\mbox{Re}\, K(z)& =& -\frac{ 4+7z} { 2z}+ \frac{2+3z} {z}\ln z-\frac{2 (1+z) ^ 2} {z ^ 2}\left(\ln
 z \ln(1+z) +\mbox{Li}_2(-z)\right), \\
\frac{1}{\pi} \mbox{Im}\, K(z) &= & - \frac{2+3z} {z} + \frac{2 (1+z) ^ 2} {z ^ 2}  \ln(1+z) .
\earray
To include a finite width, multiply the above expression by $1/(1 + i
\gamma_{G'})$ and let $m_{G'}^2 \rightarrow {m_{G'}^2 \over 1 + i
  \gamma_{G'}}$ as described above. The order $ {{g_{P}^q}^2 \over (4
  \pi)^2 }$ correction to the $Z$ width then depends on the real part
of this contribution:
\begin{multline}
\label{eq: Z partial width}
\Gamma_{Z \rightarrow q \bar{q}} = n_c {G_F m_Z^3 \over \pi 6 \sqrt{2}} \bigg( r_V(q) (f_R^q + f_L^q)^2 \big( 1 + 2 {c_F \over (4\pi)^2} { f_R^q {g_R^q}^2 + f_L^q {g_L^q}^2 \over f_R^q + f_L^q } \Re\left[ {K(m_Z^2 / M_{G'}^2) \over 1 + i \gamma_{G'}} \right] \big) \\
+ r_A(q) (f_R^q - f_L^q)^2 \big( 1 + 2 {c_F \over (4\pi)^2} { f_R^q {g_R^q}^2 - f_L^q {g_L^q}^2 \over f_R^q - f_L^q } \Re\left[ {K(m_Z^2 / M_{G'}^2) \over 1 + i \gamma_{G'}}\right] \big)  \bigg) \\
+ \Delta^q_\text{EW/QCD}
\end{multline}
where $n_c = 3$ and $r_V$ and $r_A$ are radiator factors that encode
factorizable final state QED and QCD corrections and
$\Delta^q_\text{EW/QCD}$ encodes non-factorizable corrections
\cite{Haisch:2011up}. 

For axigluon masses below $m_Z$, the $Z$ width is enhanced not only
though the vertex correction but also through real emission of an
axigluon, $Z \rightarrow q \bar{q} G'$. 
The correction to the $Z$ width from vertex corrections and from real emission of a light vector boson coupling to baryon number was computed in \cite{Carone:1994aa}, 
\beq\label{delta gamma}
\left.\frac{\Delta \Gamma (Z\to \mbox{hadrons})}{\Gamma (Z\to q\bar q)}\right|_{\mathrm{real}} =
    C \times (F_1 (x) + F_2 (x) ),
\eeq
where $C$ is a numerical constant, $F_1$ is the form factor due to real emission,
\barray
 F_1(x) &=&  (1+ x)^2 \left(3\ln x+ (\ln x) ^ 2\right) +5 (1-x ^ 2)-2x\ln x \\
       &&\phantom{move} - 2 (1+ x) ^ 2\left(\ln (1+ x)\ln x+\mbox{Li}_2 \left(\frac{1}{1+x}\right) 
        -\mbox{Li}_2 \left(\frac{x}{1+x}\right)\right) \nonumber
\earray
with $x = m_G^2/m_Z^2 = 1/z $,  and $F_2(x) = \Re[K(1/x)]$ is the form factor due to the vertex correction, which we independently computed. For flavor-universal axigluons,  in the limit as final state QED and QCD corrections are neglected, the constant in Eq.~\eqref{delta gamma} becomes $C={2 n_f c_F g_A^2 \over (4 \pi)^2}$ where $n_f = 5$. 
Because of a nontrivial cancellation of IR divergences (the limit as $x\to 0$) in the sum $F_1(x) + F_2(x)$, in the $\gamma_{G'} \rightarrow 0$ limit we can identify the form factor due to real emission of axigluons as $F_1$; 
we make the replacement $K(z) \rightarrow K(z) + F_1(1/z)$ in Eq.~\eqref{eq: Z
  partial width} to account for real emission when $m_{G'} < m_Z$. For substantial nonzero axigluon widths, $\gamma_{G'}>0$, making the replacement $K(z) \rightarrow K(z) + F_1(1/z)$ is an estimate.
  Because other issues such as the extraction of
$\alpha_s$ arise for sub-$m_Z$ axigluon masses, the estimate is sufficient for our current purposes. 

\subsection{Heavy quark contributions}\label{sec: heavy quark correction}

The off-diagonal $G'$-$q$-$Q_h$ vertex is a necessary consequence of
having quarks with phenomenogically acceptable axigluon couplings.
While the magnitude of the coupling is fixed, the mass of the heavy
quark is still a free parameter, so the minimal UV completion does not
lead to a single sharp prediction for PEW calculations.  In the
decoupling limit, $m_{Q_h}\gg m_{G'}$, the PEW calculation of the
previous subsection provides a lower bound to the total contribution.
Since the quark $Q_h$ has been proposed \cite{Tavares:2011zg} as a
possible additional decay mode to widen the axigluon, it is very
interesting to consider the cases where $2m_{Q_h}< m_{G'}$ and
$m_{Q_h} < m_{G'}$ (for a mixed $Q_h$-$q$ decay).  Specifying $\theta$
and $m_{Q_h}$ then yields a unique prediction for each pair of values
$(m_{G'}, g_P)$.

The heavy quark contributions shift the effective $Z q_P \bar{q}_P$
coupling by an amount
\beq
 \delta f_P^q =  f_P^q {{g_{\text{mix}}}^2 \over (4 \pi)^2 } c_F \,K_h (z_Z ,z_{Q_h})
\eeq
where from Eq.~\ref{eq:fulloffdiag} we have $g_{\text{mix}} = g_s \sin
2 \alpha/\sin 2 \theta $, $z_Z = m_Z^2/ m_{G'}^2$, $z_{Q_h} =
m_{Q_h}^2/ m_{G'}^2 $, and the form factor $K_h(z_Z, z_{Q_h})$ is given by the
following integral over Feynman parameters,
\beq
K_h(z_Z,z_{Q_h}) = \int_0^1 \int_0^{1-x} f_\text{1}(x,y; z_Z, z_{Q_h}) \,dy\,dx + \int_0^1 f_\text{2}(x; z_{Q_h}) \, dx
\eeq
where
\beq
f_2(x; z_{Q_h}) = 2 - 2 x^2 (1 - z_{Q_h}) + (1 + 3 x^2(1-z_{Q_h}) - x(4+z_{Q_h})) \log(1 - x (1 - z_{Q_h})).
\eeq
and
\begin{multline}
f_\text{1}(x,y; z_Z, z_{Q_h}) = - \Delta_\text{1} + { ((1 - x) (1 - y) z_Z + z_{Q_h}) (2 + x y z_Z) \over \Delta_1 } -  (4 - z_Z (x+y - 2 x y) + z_{Q_h}) \\
+(4 - z_Z  - z_{Q_h} + 3 ( x + y) (z_Z -2(1 - z_{Q_h}) )  
  - 12 x y z_Z) \log(\Delta_1),
\end{multline}
with $ \Delta_\text{1} = 1 - x y z_Z + (x+y) (z_{Q_h} - 1)$.

\begin{figure}
\includegraphics[width=0.48 \textwidth]{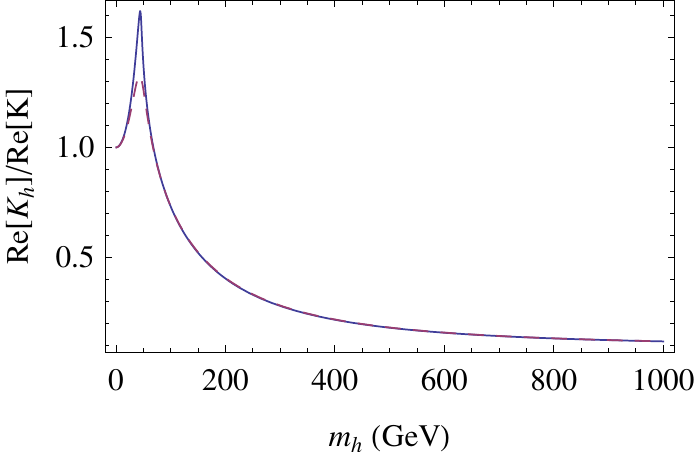}
\includegraphics[width=0.48 \textwidth]{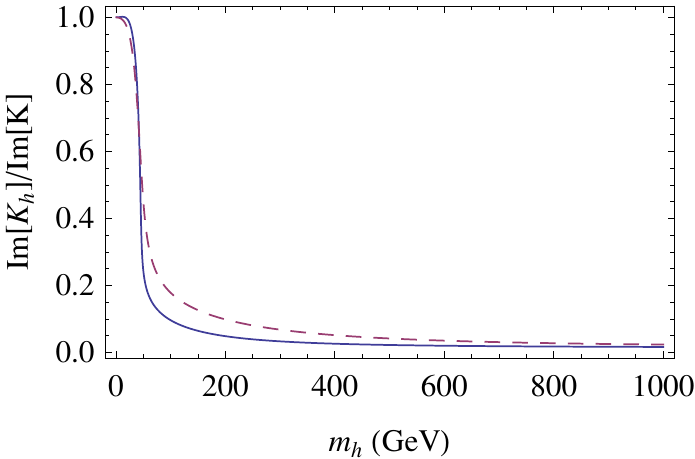}
\caption{The ratios $\Re [ K_h\left( z_Z, z_{Q_h} \right) ] /  \Re [ K(z_Z) ] $ and $\Im [ K_h\left( z_Z, z_{Q_h} \right) ] /  \Im [ K(z_Z) ] $ are plotted for $z_Z = {m_Z^2(1 + i \gamma_{G'}) \over m_{G'}^2 (1 + i \gamma_{Z})}$, $z_{Q_h} = { m_{h}^2 (1 + i \gamma_{G'})  \over m_{G'}^2 (1 + i \gamma_{h}) } $, with $m_Z = 91.2$ GeV, $\gamma_Z = 2.50 / 91.2$, $m_{G'} = 250$ GeV, $\gamma_{G'} = 0.2$ and $\gamma_h = 0.1$ (blue, solid) and $0.4$ (pink, dashed).}\label{fig: heavy quarks}
\end{figure}

In the limit as $m_{Q_h} \rightarrow 0$, $K_h$ reduces to $K$ in
Eq.~\eqref{eq: vertex correction}: $K_h(z_Z, 0) = K(z_Z)$. In the
limit as $m_{Q_h} \rightarrow \infty$, $K_h \rightarrow (7/36) m_Z^2 /
m_{G'}^2$. Note that although $K_h$ is finite in the decoupling limit,
the overall contribution of the heavy fermion still decouples, as the
prefactor contains the coupling $g^2_{mix}$, which scales like
$m_{Q_h} ^ {-2}$ as $m_{Q_h} \rightarrow \infty$ with $\lambda$
fixed. The ratios $\Re [ K_h\left( z_Z, z_{Q_h} \right) ] / \Re [
K(z_Z) ] $ and $\Im [ K_h\left( z_Z, z_{Q_h} \right) ] / \Im [ K(z_Z)
] $ are plotted in Fig.~\ref{fig: heavy quarks}.  Note that the sign
of these contributions is the same as that of the contribution from
the axigluon alone, and thus including these contributions to the $Z$
vertex correction will also act uniformly to increase the effective
$Z$-$q$-$\bar q $ coupling. Therefore including heavy quarks as
additional decay modes for the axigluon only increases the constraints
from PEW observables.

In general, there will also be contributions to the effective coupling
from the uneaten part of the field that breaks $SU(3)_1\times
SU(3)_2\to SU(3)_c$ via Eq.~\eqref{eq:lmix}. The scalar-heavy-quark
contributions shift the $Z q_P \bar{q}_P$ coupling by
\beq
 \delta f_P^q =  f_P^q {{\lambda^2 \cos^2 \alpha } \over 6 (4 \pi)^2 }  \,K_\phi \left( {m_Z^2 \over m_{\phi}^2} , {  m_{Q_h}^2 \over m_{\phi}^2}\right).
\eeq

Here,
\beq\label{eq: gmix scalar relationship}
{\lambda^2 \cos^2 \alpha \over 6} =  c_F g_\text{mix}^2 {1 \over 8}{m_{Q_h}^2 \over m_{G'}^2}
\eeq
 so
for heavy quark masses less than $2 \sqrt{2}/g_s$ times the axigluon mass,
the coefficient entering the scalar-heavy quark correction is less
than that entering the heavy quark-axigluon correction.

Let $z_{Z} = {m_Z^2 \over m_\phi^2}$ and $z_{Q_h}={m_{Q_h}^2 \over
 m_{\phi}^2}$. The form factor $K_\phi$ is given by the following
integral over Feynman parameters,
\beq
K_\phi (z_Z, z_{Q_h} ) = \int_0^1 \int_0^{1-x} f_\text{3}(x,y; z_Z, z_{Q_h}) \,dy\,dx + \int_0^1 f_\text{4}(x; z_{Q_h}) \, dx
\eeq
where
\beq
f_4(x; z_{Q_h}) = -x \log( x+ z_{Q_h} (1-x))
\eeq
and
\beq
f_3(x,y; z_Z, z_{Q_h}) =1 - { x y z_Z + z_{Q_h} \over \Delta_3} + \log(\Delta_3),
\eeq 
with $\Delta_3 = (x+y) z_{Q_h}  + (1-x-y) - x y z_Z $.

We find the following limiting behavior of $K_\phi$:
\beq
K_\phi \longrightarrow \Bigg \lbrace 
\begin{array}{c l}
{1 \over 3} {m_Z^2 - 3 m_{Q_h}^2 \over m_{\phi}^2}  & \text{~~as~~} m_{\phi} \rightarrow \infty \\
-{7 \over 36} {m_Z^2 \over m_{Q_h}^2} & \text{~~as~~} m_{Q_h} \rightarrow \infty \\
f_0(m_Z^2 / m_{\phi}^2) & \text{~~as~~} m_{Q_h} \rightarrow 0 
\end{array}
\eeq
where 
\beq
f_0(x) = {2 \log x - 1 \over 4} + {1-\log x \over x} + {\log x \log (1+x) + \text{Li}_2(-x) \over x^2} - i \pi \left( {\log(1+x) \over x^2} - {1 \over x} + {1 \over 2} \right).
\eeq

\begin{figure}
\includegraphics[width=0.48 \textwidth]{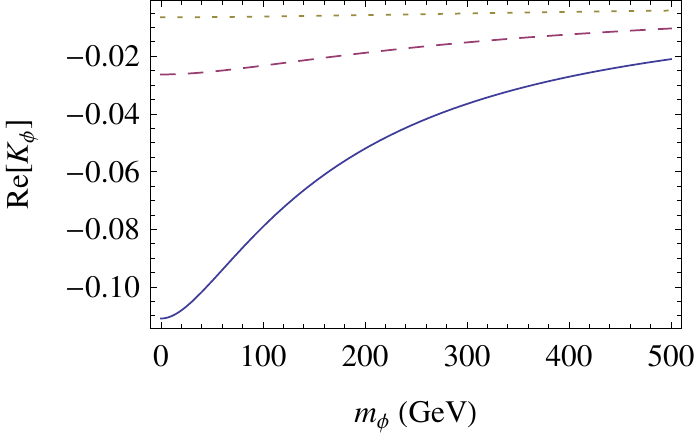}
\includegraphics[width=0.48 \textwidth]{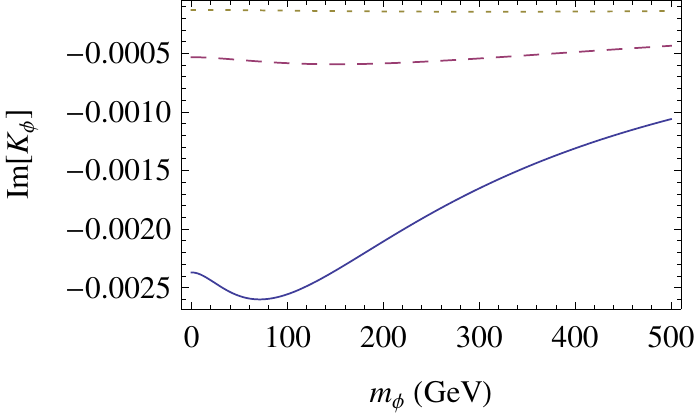}
\caption{$\Re [ K_\phi \left( z_Z, z_{Q_h} \right) ] $ and $\Im [ K_\phi \left( z_Z, z_{Q_h} \right) ] $ are plotted for $z_Z = {m_Z^2(1 + i 0.1) \over m_{\phi}^2 (1 + i 0.03) }$, $z_{Q_h} = { m_{Q_h}^2 (1 + i 0.1) \over m_{\phi}^2 (1+i 0.05) } $, with $m_Z = 91.2$ GeV, $m_{Q_h} = 125$ GeV (solid blue), 250 GeV (dashed pink), and 500 GeV (dotted yellow).  }\label{fig: scalar correction}
\end{figure}

\begin{figure}
\includegraphics[width=0.48 \textwidth]{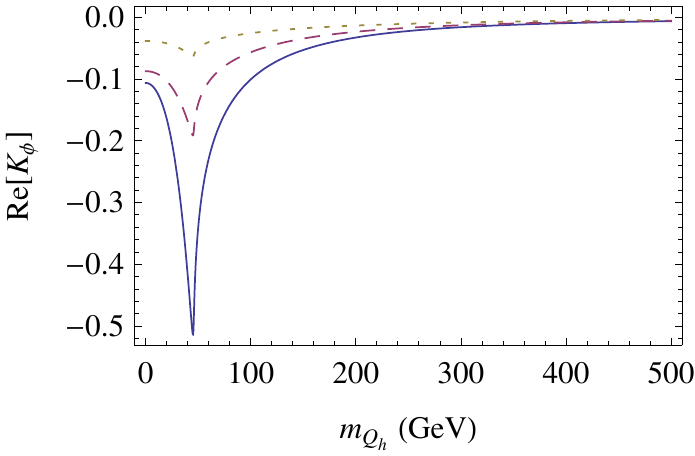}
\includegraphics[width=0.48 \textwidth]{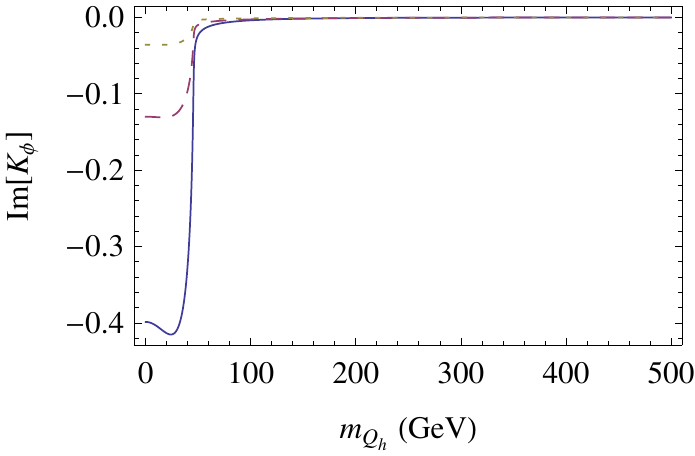}
\caption{$\Re [ K_\phi \left( z_Z, z_{Q_h} \right) ] $ and $\Im [ K_\phi \left( z_Z, z_{Q_h} \right) ] $ are plotted for $z_Z = {m_Z^2(1 + i 0.1) \over m_{\phi}^2 (1 + i 0.03) }$, $z_{Q_h} = { m_{Q_h}^2 (1 + i 0.1) \over m_{\phi}^2 (1+i 0.05) } $, with $m_Z = 91.2$ GeV, $m_{\phi} = 125$ GeV (solid blue), 250 GeV (dashed pink), and 500 GeV (dotted yellow).  }\label{fig: scalar correction 2}
\end{figure}

We find that the scalar contribution has the opposite sign as the
heavy quark-axigluon contribution, which could serve to moderate
precision electroweak constraints for certain regions of parameter
space. In Fig.~\ref{fig: scalar correction} we plot $\Re K_\phi$ and $\Im K_\phi$ as a function of $m_\phi$ assuming $m_{Q_h} = 125$ GeV (solid blue), 250 GeV (dashed pink), and 500 GeV (dotted yellow). By comparison, $\Re K(m_Z^2 / (250 \text{GeV})^2) = 0.33$ and $\Re K(m_Z^2 / (100 \text{GeV})^2) = 1.00$. In Fig.~\ref{fig: scalar correction 2} we plot the real and imaginary contributions as functions of $m_{Q_h}$, with  $m_\phi$ held fixed.

\end{document}